\documentclass[fleqn,usenatbib]{mnras}
\usepackage{subcaption}
\usepackage{booktabs}
\usepackage{float}
\usepackage{soul}
\usepackage{newtxtext,newtxmath}
\usepackage[normalem]{ulem}

\usepackage[T1]{fontenc}
\usepackage{multirow}

\usepackage{graphicx}	
\usepackage{amsmath}	
\usepackage{color}
\usepackage{natbib}
\usepackage{url}
\usepackage{placeins}
\usepackage{subcaption}
\usepackage{xfrac}
\usepackage{comment}

\definecolor{gian}{rgb}{1.0, 0., 1.0}

\definecolor{amethyst}{rgb}{0.8, 0.0, 0.0}

\definecolor{orange}{rgb}{1, 0.5, 0.}

\definecolor{green}{rgb}{0., 0.75, 0.}
\newcommand{\rev}[1]{{#1}}
\DeclareRobustCommand{\VAN}[3]{#2}
\let\VANthebibliography\thebibliography
\def\thebibliography{\DeclareRobustCommand{\VAN}[3]{##3}\VANthebibliography}

\usepackage{graphicx}	
\usepackage{amsmath}	

\title[The $z \lesssim 1$ drop of cosmic dust in a SAM framework]{The $z \lesssim 1$ drop of cosmic dust abundance in a semi-analytic framework}

\author[M. Parente et al.\ ]{Massimiliano Parente $^{1,2}$\thanks{E-mail: mparente@sissa.it}, Cinthia Ragone-Figueroa$^{3,2}$, Gian Luigi Granato$^{2,3,4}$, Andrea Lapi$^{1,4,5,6}$
\\
\\
$^{1}$ SISSA, Via Bonomea 265, I-34136 Trieste, Italy \\
$^{2}$ INAF, Osservatorio Astronomico di Trieste, via Tiepolo 11, I-34131, Trieste, Italy \\
$^{3}$ Instituto de Astronom\'ia Te\'orica y Experimental (IATE), Consejo Nacional de Investigaciones Cient\'ificas y T\'ecnicas de la\\ Rep\'ublica Argentina (CONICET), Universidad Nacional de C\'ordoba, Laprida 854, X5000BGR, C\'ordoba, Argentina\\
$^{4}$ IFPU - Institute for Fundamental Physics of the Universe, Via Beirut 2, 34014 Trieste, Italy\\
$^{5}$ INFN - National Institute for Nuclear Physics, Via Valerio 2, I-34127 Trieste, Italy\\
$^{6}$ INAF - Istituto di Radio-Astronomia, Via Gobetti 101, I-40129 Bologna, Italy\\
}

\date{Accepted XXX. Received YYY; in original form ZZZ}

\pubyear{2022}

\begin{document}
\label{firstpage}
\pagerange{\pageref{firstpage}--\pageref{lastpage}}
\maketitle

\begin{abstract}
Observations suggest that the amount of galactic dust in the Universe decreased by a factor $\sim 2-3$ during the last $\sim 8$ Gyr. However, cosmological models of galaxy evolution usually struggle to explain this decrease. Here we use the semi-analytic model (SAM) \textsc{L-Galaxies2020} to show that this drop may be reproduced assuming standard prescriptions for dust production and evolution. We extend the SAM with \textit{i)} a state-of-the-art dust model which adopts the two-size approximation and \textit{ii)} a new disc instability criterion which triggers bulge and central black hole growth.
The model reproduces some fundamental properties of the local galaxy population, such as the fraction of spheroid-dominated galaxies and some scaling relations involving dust. Moreover, the model predicts a galactic dust drop from $z \sim 1 \rightarrow 0$, which becomes closer to the observed one when adopting the new treatment of disc instabilities. This result is related to the newly implemented super-massive black hole growth during disc instabilities, which enhances the quenching of massive galaxies. Consequently, these objects feature a lower gas and dust content.
We provide a census of the contribution of all the processes affecting the galactic dust content. Accretion is the dominant dust mass growth process. Destruction by supernovae, astration and ejection by winds have all a non-negligible role in decreasing the overall dust content in galaxies below $z \sim 1$. We also discuss predictions concerning extra-galactic dust, confirming that a sputtering efficiency lower than the canonical one is required to match the few available observations.
\end{abstract}

\begin{keywords}
galaxies: evolution –- galaxies: formation -– galaxies: ISM –- galaxies: general –-  ISM: dust
\end{keywords}



\section{Introduction}

Dust grains are small ($\sim 1-10^{-3}\, \mu{\rm m}$) solid particles made of heavy metals polluting the interstellar medium (ISM) of galaxies (see \citealt{Galliano2018} for a recent review). The presence of such grains heavily affects the observed spectral energy distribution (SED) of galaxies, since they absorb UV radiation and re-emit it in the IR region.
These effects strongly depend not only on grain properties but also on the geometry \citep[e.g.][]{Granato2000}. In addition, dust plays a vital role in galaxy evolution, since it actively participates in many physical processes. For example, cooling of the hot gas may be contributed by ions-grains collisions \citep[e.g.][]{Burke1974,Dwek1981,Montier2004}. The formation of H$_2$ molecules occurs predominantly on the surface of dust grains \citep{Wakelam2017}, which are also able to shield molecular clouds from LW dissociating radiation. Consequently, grains play a crucial role in the whole star formation process. Moreover, radiation pressure on dust grains may stimulate the development of galactic winds (e.g. \citealt{Murray2005}). Finally, radiation drag on dust grains may promote the accretion of low angular momentum gas to the central supermassive black hole (SMBH) of galaxies \citep[e.g.][]{Granato2004}.\\

Given its relevance, a detailed understanding of the build-up of dust in galaxies across cosmic time would be highly advantageous. In brief, dust grains are produced in stellar ejecta due to the condensation of  a certain fraction of some heavy element. Asymptotic Giant Branch stars (AGBs) winds and Supernovae (SNe) explosions are currently believed to be the leading factories of dust \citep[e.g.][]{Gail2009, Bianchi2007}. However, once ejected into the ISM, several processes heavily modify the properties of grains \citep[e.g.][]{Dwek1998, Zhukovska2008}. In dense regions, gas-phase metals may accrete onto grains, thus increasing the dust mass. On the other hand, grains may be eroded (\textit{sputtered}) in SN shocks and by collisions with highly energetic ions in the hot plasma.
Collisions between grains affect their size distribution: \textit{coagulation} in dense environment, resulting from low velocity ($\lesssim 0.1-1$ km/s) collisions, and \textit{shattering} in the diffuse medium, produced by high velocity collisions, act to shift the distribution towards large and small sizes, respectively, without affecting directly the dust mass. However, accretion and sputtering, being surface processes, are more effective on smaller grains. As a result, the mass, chemical composition and size distribution of grains is a product of the interplay between the various processes, which in turn depend on the variety of physical conditions in the ISM (e.g. \citealt{Aoyama2017}; \citealt{Parente2022}).\\

All the aforementioned processes, whose relative efficiency is still uncertain and debated \citep[e.g.][]{Ferrara2016,Vijayan2019,Triani2020,Dayal2022}, contribute to the total budget of dust in our Universe. This quantity, commonly given in terms of $\Omega_{\rm dust}=\rho_{\rm dust}/\rho_{\rm c,0}$\footnote{Here $\rho_{\rm dust}$ is the comoving dust mass density, and $\rho_{\rm c,0}=2.775 \,h^2 \times 10^{11} \, M_\odot/{\rm Mpc}^3$ is the critical density of the Universe today.}, can be thought as the sum of a galactic and an extra-galactic component. In this work we refer to them as $\Omega^{\rm ISM}_{\rm dust}$ and $\Omega^{\rm CGM}_{\rm dust}$  respectively. In the last two decades, observations have made possible the determination of both of them, albeit with unavoidable uncertainties. 

The cosmic abundance of galactic dust $(\Omega^{\rm ISM}_{\rm dust})$ may be estimated by deriving the dust mass of a sample of galaxies by fitting their SED, then building a Dust Mass Function (DMF) and integrating it.
Different examples exist in the literature. \cite{Dunne2011} exploited a sample of $\sim 2000$ sources from the Herschel-ATLAS survey, each of which with a reliable counterpart from the Sloan Digital Sky Survey (SDSS), constraining the evolution of the DMF at $z < 0.5$. They found that high$-z$ massive galaxies have more dust than local galaxies, and that $\Omega^{\rm ISM}_{\rm dust}$ decreases with decreasing redshift. \cite{Driver2018} derived dust masses of $\sim 570000$ sources from the GAMA, G10-COSMOS, and 3D-HST catalogues over a broad redshift range ($0 < z \lesssim 2$), finding as well a smooth decline of $\Omega^{\rm ISM}_{\rm dust}$ at $z \lesssim 1$. More recently, \cite{Pozzi2020} largely improved our understanding of the cosmic evolution of the DMF, by studying it up to $z \simeq 2.5$. Their results broadly confirmed what was found in past works, that is a broad peak of $\Omega^{\rm ISM}_{\rm dust}$ at $z \sim 1$, and then a decline toward the local Universe. Also $z \sim 0$ determinations of the DMF are available today, and are consistent with the findings highlighted above. To mention one representative work, \cite{Beeston2018} studied the DMF of a sample of $z < 0.1$ galaxies from the Herschel-ATLAS and GAMA surveys, putting precise constraints on $\Omega^{\rm ISM}_{\rm dust} (z\simeq0)$.

By converse, measurements of the
amount of extra-galactic dust ($\Omega^{\rm CGM}_{\rm dust}$) are more complex and, as a consequence,  more scanty and uncertain. Remarkably, \cite{Menard2010} (see also \citealt{Menard2012}; \citealt{Peek2015}) derived an estimate of dust in galactic halos\footnote{\rev{However, some of the absorbers selected by \cite{Menard2012} feature equivalent widths which are associated with column density typical of neutral gas, thus ISM (see $\S 4.2$  of \citealt{Peroux2020}).}} by exploiting the reddening of distant quasars by foreground absorbers, showing that the amount of dust residing beyond galaxies is not negligible. \\

The result emerging from these observations, when taken as a whole (e.g. \citealt{Peroux2020}, their Fig. 12), is that the abundance of dust in galaxies has decreased by a factor $\sim 2-3$ during the last $\sim 8 \, \text{Gyr}$. This behaviour is still not clear from a theoretical point of view. Indeed, the relatively recent inclusion of dust in some semianalytic-models (\citealt{Popping2017}; \citealt{Vijayan2019}; \citealt{Triani2020}) and hydrodynamic cosmological simulations (\citealt{McKinnon2017}; \citealt{Aoyama2018}; \citealt{Hou2019}; \citealt{Li2019}; \citealt{Graziani2020}; \citealt{Parente2022}) of galaxy evolution, has made it possible to study the build-up of the DMF across cosmic epochs, investigating the role of the various processes shaping dust evolution in a cosmological context. However, most of the aforementioned works do not reproduce the observed peak of $\Omega^{\rm ISM}_{\rm dust}$. Remarkable exceptions are the hydrodynamic simulations carried out by \cite{Aoyama2018} and \cite{Li2019}. The first group reproduces a decline of $\Omega^{\rm ISM}_{\rm dust}$ at $z <1$, attributing it to astration, but their normalization is a factor $\sim 6$ too high at $z=0$. The second group provides an excellent fit of the observed galactic $\Omega^{\rm ISM}_{\rm dust}$ across cosmic time. However, they do not discuss the origin of its behaviour. \\

Motivated by this issue, \cite{Ferrara2021} tried to put some constraints on the efficiency of processes acting to destroy dust, assuming that they overcome\footnote{This is needed to obtain a decline of $\Omega_{\rm dust}$, since they consider both galactic and extra-galactic dust.} the processes responsible for dust growth at $z<1$.  
However, their simple computation only takes into account the efficiency of dust processes for $z < 1$, neglecting the impact of these processes on the previous evolution of $\Omega_{\rm dust}$. \\

In this work, we aim to study the $\Omega^{\rm ISM}_{\rm dust}$ peak at $z \simeq 1$ and its later decline in a semi-analytic framework. We focus mainly on the contribution to $\Omega_{\rm dust}$ coming from dust inside galaxies, which is observed to drop at $z\lesssim1$, and it is better constrained by observations. However, we also briefly discuss current observations concerning dust outside galaxies. Our goal is to assess the relative importance of the aforementioned dust-related processes in determining the decrease. A semi-analytic approach is particularly suitable for this kind of study since it allows a relatively computationally cheap exploration of the role of such processes, still considering several baryonic processes shaping galaxy evolution which may as well have a profound impact on $\Omega_{\rm dust}$. For this reason, we exploit the latest public release of the code \textsc{L-Galaxies}2020 \citep{Henriques2020} and extend it with a detailed treatment of dust formation and evolution in galaxies. We anticipate here that we also modify the SAM treatment of disc instabilities: besides improving the resulting morphology of the simulated galaxies, the growth of SMBH during instabilities and the consequently induced quenching turns out to be of fundamental importance for reproducing a $\Omega^{\rm ISM}_{\rm dust}$ consistent with observations.

The paper is organized as follows. We recap the main features of the SAM in Sec. \ref{sec:sam}, where we also detail the new implementation of dust evolution and disc instabilities (Sec. \ref{sec:dustmodel} and \ref{sec:DImodel}). We present and discuss our results in Sec. \ref{sec:results}. Here we first check the consistency between our model predictions and some crucial quantities (Sec. \ref{sec:sanity} and \ref{sec:morphoSF}), then we show our results concerning dust in Sec. \ref{sec:dust_scaling}. The evolution of $\Omega_{\rm dust}$ in both galactic and extra-galactic environments is discussed at length in Sec. \ref{sec:dust_cosmic}. We summarize our work and present the conclusions in Sec. \ref{sec:conclusion}.

\section{The semi-analytic model}
\label{sec:sam}

\begin{table}
 \begin{tabular}{c c c} 
 \hline
 \hline
 {} & \multicolumn{2}{c}{Reference in \cite{Henriques2020}} \\
 {Physical Process} & \multicolumn{2}{c}{supplementary material}\\
 {} & {Equations} & {Sections} \\[0.5ex]
\hline\hline
Gas infall into DM halos & S2 & S1.3 \\ \midrule[0.2\arrayrulewidth]
Gas cooling into gaseous disc & S6, S7 & S1.4 \\ \midrule[0.2\arrayrulewidth]
Spatially resolved properties & \multirow{2}{*}{S8$-$S11, S18} & \multirow{2}{*}{S1.5, S1.6, S1.9} \\ 
of discs & & \\
\midrule[0.2\arrayrulewidth]
H$_2$-based star formation & S12$-$S16 & S1.7, S1.8 \\ \midrule[0.2\arrayrulewidth]
SNe feedback, gas ejection & \multirow{2}{*}{S19, S23, S24} & \multirow{2}{*}{S1.10, S1.11} \\
and reincorporation & & \\ \midrule[0.2\arrayrulewidth]
\multirow{2}{*}{Chemical enrichment} & \multirow{2}{*}{S26} & {S1.13} \\
{} & {} & and \cite{Yates2013} \\ \midrule[0.2\arrayrulewidth]
SMBHs growth & \multirow{2}{*}{S27$-$S29} & \multirow{2}{*}{S1.14} \\
and feedback & & \\ \midrule[0.2\arrayrulewidth]
Environmental processes & \multirow{2}{*}{S31, S33, S34} & \multirow{2}{*}{S1.15} \\
and satellites distruption & & \\ \midrule[0.2\arrayrulewidth]
Mergers and & \multirow{2}{*}{S36, S37} & \multirow{2}{*}{S1.16.1, S1.16.2}\\ 
starbursts & & \\ \midrule[0.2\arrayrulewidth]
Bulges and & \multirow{2}{*}{S38, S40} & \multirow{2}{*}{S1.16.3}\\
disc instabilities & & \\

\hline\hline
\end{tabular}
\caption{\rev{List of the main physical processes modeled in the \textsc{L-Galaxies 2020} SAM. For each of them, we refer to the appropriate equations and sections of the supplementary material of \citet{Henriques2020} describing the processes and their relative implementation.}}
\label{tab:SAMoverview}
\end{table}
\rev{In general, SAMs apply to DM haloes merger trees approximate descriptions for the baryonic processes which are believed to be relevant in shaping galaxy populations. The merger tree can be either extracted from gravity-only simulations or calculated by Monte Carlo methods. The former possibility has become more common in the last decade. These baryonic processes now routinely include gas infall into DM halos, gas cooling, star formation, chemical and energetic feedback (the latter both from stars and AGNs), gravitational instabilities and interactions of galaxies with the environment. The {\it assumed} picture is that the first outcome of gas collapse is the formation of gas disks, supported by rotation and featuring a mild star formation activity. The spheroidal component of galaxies results from galaxy mergers and instabilities, which can also produce violent starbursts if enough gas is present. The processes are described by approximate relationships between some galaxy properties (mostly mass budget in a few components and scale lengths), which are used to evolve the galaxy population over timesteps.}
In particular, here we adopt the public\footnote{The source code is available at \url{https://github.com/LGalaxiesPublicRelease/LGalaxies_PublicRepository/releases/tag/Henriques2020}.} semi-analytic model of galaxy evolution \textsc{L-Galaxies 2020} (\citealt{Henriques2020}), the latest release of the Munich galaxy formation model. This SAM is designed to run on the Dark Matter (DM) merger trees of the \textsc{Millennium} and \textsc{Millennium-II} simulations (\citealt{Springel2005}; \citealt{Boylan-Kolchin2009}), 
\rev{We refer the reader to the \textsc{L-Galaxies} papers (\citealt{Henriques2015}; \citealt{Henriques2020}), and to their supplementary material available online at {\url{https://lgalaxiespublicrelease.github.io/Hen20_doc.pdf}} for a complete description of the SAM. We list in Tab.\ref{tab:SAMoverview}
the physical processes treated by the SAM prescriptions, and the relevant sections and equations.}

\rev{In the \textsc{L-Galaxies2020} model, there are} a few relevant differences with respect to the previous version presented by \cite{Henriques2015}. First, galactic discs (both gas and stars) are spatially resolved to some extent, i.e. they are divided into $12$ concentric rings with radii $r_i = 0.01 \cdot 2^i h^{-1} \, {\rm kpc}$ ($i=0,...,11$) (\citealt{Fu2013}). Therefore, all disc properties and processes are followed for each ring separately (e.g. star formation, chemical enrichment, gas ejection). In particular, star formation is linked to the H$_2$ amount of each ring, which is in turn modeled according to a metallicity-dependent description (\citealt{Krumholz2009}; \citealt{McKee2009}). Of \rev{particular} relevance here is the inclusion of the galactic chemical enrichment (GCE) model of \cite{Yates2013}, which allows tracking the amount of 11 elements, released to the gas phase by AGB stars, SNIa and SNII.

Finally, the free parameters of the SAM have been calibrated by means of a MCMC sampling technique (\citealt{Henriques2009,Henriques2015}) in order to fit a number of observational constraints at $z=0$ and $z=2$ which we left unchanged.\\

We run the model on the \textsc{Millennium} merger trees, (original box size $500/h$, $2160^3$ particles)  adopting a \textit{Planck} cosmology\footnote{The original \textsc{Millennium} cosmology has been scaled according to \cite{Angulo2010} and \cite{Angulo2015}.} (\citealt{Planck14}) with $h=0.673$, $\Omega_{\rm m}= 0.315$, $\Omega_{\rm b}= 0.0487$, $\sigma_8=0.829$. We adopt a \cite{Chabrier2003} initial mass function (IMF). In selecting galaxies for our analysis we consider both central and satellite galaxies with ${\rm log} M_*/M_\odot > 8.5$ and ${\rm log} M_{\rm HI}/M_\odot > 8$.

\subsection{Dust production and evolution}
\label{sec:dustmodel}
The dust formation and evolution model implemented in this work is conceptually similar to that implemented recently by our group (\citealt{Gjergo2018}; \citealt{Granato2021}; \citealt{Parente2022}) in hydrodynamic simulations. However, here we adapt it to the semi-analytic framework of \textsc{L-Galaxies}. As in the aforementioned works, we aim to follow the mass evolution of two populations of grains with different sizes, which we will refer to as \textit{large} and \textit{small} grains. Indeed, we adopt the two-size approximation proposed by \cite{Hirashita2015}, who showed it to be a numerically cheap method to capture reasonably well the effects of the various processes considered in dust evolution models including grain size distribution. The method has been further validated by \cite{Aoyama2020}. Moreover, we independently follow carbonaceous and silicate dust grains.
The four species pollute the cold gas, hot gas, and ejected reservoir of galaxies, where they are subjected to different processes affecting their abundance. 

The aim of this section is to detail the implementation within the SAM of these processes: production by stellar sources (Sec. \ref{sec:dustprod}), shattering and coagulation (Sec. \ref{sec:dustshatcoa}), grain growth in molecular clouds (Sec. \ref{sec:dustgrowth}), destruction in SNe shocks (Sec. \ref{sec:dustSNshock}), sputtering in the hot phase and ejected reservoir (Sec. \ref{sec:dustspu}). \   \
For the sake of clarity, we anticipate here that at each timestep of the simulation we compute the mass rate of each process and then use it to evaluate the overall mass variation of small ($_{\rm S}$) and large ($_{\rm L}$) grains in the cold gas, hot gas, and ejected reservoir of a galaxy:
\begin{align}
     \dot{M}_{\rm L}^{\rm cold} = &\dot{M}_{\rm *}^{\rm cold} + \dot{M}_{\rm coag} - \dot{M}_{\rm shat} - \dot{M}_{\rm astr} - \dot{M}_{\rm SN} ,  \\
    \dot{M}_{\rm S}^{\rm cold} = &\dot{M}_{\rm acc} - \dot{M}_{\rm coag} + \dot{M}_{\rm shat}  - \dot{M}_{\rm astr} - \dot{M}_{\rm SN}, \\
     \dot{M}_{\rm L}^{\rm hot} = &\dot{M}_{\rm *}^{\rm hot} - \dot{M}_{\rm spu}, \\
    \dot{M}_{\rm S}^{\rm hot} = & - \dot{M}_{\rm spu},  \\
     \dot{M}_{\rm L,S}^{\rm ej}  = & - \dot{M}_{\rm spu}.
\end{align}

In the above equations, $\dot{M}_{*}$ is the dust produced by stars, which are assumed to enrich both the cold and hot gas with large grains, $\dot{M}_{\rm acc}$ is the dust grown by metal accretion, acting on small grains only. The $\dot{M}_{\rm shat}$ and $\dot{M}_{\rm coag}$ terms describe the shattering and coagulation processes, which exchange mass between small and large grains, without modifying the total dust budget.  $\dot{M}_{\rm astr}$ is the dust removed from the cold gas phase during star formation by astration, and $\dot{M}_{\rm SN}$ is the dust destroyed in SN shocks. $\dot{M}_{\rm spu}$ is the term describing thermal sputtering, which we model for dust in both the hot gas and ejected reservoir\footnote{\label{ejfootnote}The ejected reservoir is an extra-galactic component made of hot gas that can not cool. It is enriched by SN driven winds and such material may be reincorporated into the hot gas, making it available again for cooling. The ejection/reincorporation prescriptions adopted by the SAM (see supplementary material of \citealt{Henriques2020}) are tailored to stop star formation for relatively long periods in low mass systems. The position of the ejected reservoir is not specified, so its physical interpretation is ambiguous (see Sec. \ref{sec:Omega_hot}).}.
In the following, we give a detailed description of the processes mentioned above. Most of them are modeled by computing an associated timescale $\tau_{\rm process}$, from which the mass rate is:
\begin{equation}
    \dot{M}^{\rm cold, hot, ej}_{\rm S,L} = \frac{M^{\rm cold, hot, ej}_{\rm S,L}}{\tau_{\rm process}}.
\end{equation}

We note that in all the processes described by the SAM which cause gas mass transfer between components (e.g. in the ejection of cold gas by SN-driven winds, in cooling from hot to cold gas, in galaxy mergers, etc.), we generally assume that dust is preserved. These variations of dust content  are not described by the above equations, where we instead highlight astration, since, in this case, dust is immediately destroyed and not just moved from one component to another.

\subsubsection{Stellar production}
\label{sec:dustprod}
Following \cite{Hirashita2015}, stellar populations enrich their surrounding medium with gas metals and large dust grains. In particular, winds of AGB stars and SNe ejecta are expected to be the main production sources of grains. In this work, we rely on the chemical enrichment model adopted by the SAM \citep{Yates2013}, which allows tracking the chemical enrichment of eleven individual elements from SNII, SNIa, and AGB stars. We assume that a certain fraction of these elements, specified in the following, condense into dust grains. Before going into details, we anticipate a few features of our implementation.
\begin{itemize}
    \item[(i)] We assume that dust may be produced only by AGB stars and in SNII ejecta, thus we neglect the contribution of SNIa (see e.g. \citealt{Gioannini2017}; \citealt{Li2019}; \citealt{Parente2022}).
    \item[(ii)] Stellar populations enrich with metals and dust both cold and hot gas, according to the assumptions of the SAM\footnote{See also \cite{Yates2021} for a discussion on the direct hot phase metal enrichment by stars.}. Namely, disk SNII have an enrichment efficiency of the hot gas $f_{\rm SNII, hot}=0.3$, while for AGB stars $f_{\rm AGB, hot}=0.0$. Bulge and ICL stars enrich only the hot medium\footnote{This hot gas dust enrichment due to bulge and ICL stars has negligible impact on our results: our findings are almost unchanged when assuming that dust produced by bulge and ICL stars is destroyed when released into the hot gas. The main contribution to the dust enrichment of hot gas comes from stellar driven winds, which move dust from the cold to the hot gas.}. 
    \item[(iii)] We consider two chemical compositions of dust: carbonaceous and silicate grains. The former is made only of C atoms, while we assume an olivine-like composition MgFeSiO$_4$ for silicates. \\
    
\end{itemize}

AGB stars are assumed to form carbonaceous or silicate grains, depending on the C/O number ratio in the ejecta. Indeed, since the ejecta are mixed at the microscopic level, the maximum possible amount of CO is formed before grain condensation, leaving available only the remaining C or O atoms \citep[e.g.][]{Dwek1998}. When ${\rm C/O}>1$, AGB stars produce carbon dust using the C atoms not locked into CO:
\begin{equation}
    M_{\rm C \, dust} = \text{max}\left[ \delta_{\rm AGB,C} \left( M_{\rm C\, ej} - 0.75 M_{\rm O \, ej}\right), 0 \right],
\end{equation}
where $M_{\rm X \, ej}$ is the ejected mass of the ${\rm X}$ element and 0.75 is the ratio between O and C atomic weights.
We set the condensation efficiency $\delta_{\rm AGB,C}=0.1$. 

AGB stars produce silicates when ${\rm C/O}<1$. In order to preserve the chemical composition of Olivine, we first evaluate the element which constitutes a \textit{bottleneck} for its formation. This element, often referred to as \textit{key element} (see e.g. \citealt{Zhukovska2008}), is the one that minimizes the ratio between the number of atoms ejected and the number of atoms entering the compound. This value sets the number of units of MgSiFeO$_4$ that may form in the ejecta, that is:
\begin{equation}
\label{eq:Nsil}
    N_{\rm sil} = \delta_{\rm AGB, sil} \min_{X \in \text{[Mg,Fe,Si,O]}} \left(\frac{M_{{\rm X}\, \rm ej}}{\mu_{\rm X} N^{\rm X}_{\rm ato}}\right),
\end{equation}
where $\mu_{\rm X}$ is the atomic weight of the element, $N^{\rm X}_{\rm ato}$ the number of ${\rm X}$ atoms in the compound, and $\delta_{\rm AGB,sil}=0.1$. Once $N_{\rm sil}$ is computed, the mass of each element ${\rm X} \in \text{[Mg,Fe,Si,O]}$ condensing into dust is simply given by:
\begin{equation}
    M_{ {\rm X}\, \rm dust} = N_{\rm sil} \mu_{\rm X} N_{\rm ato}^X.
\end{equation}
\\
Differently from AGB stars, in SNII ejecta the formation of carbon and silicate dust grains are not mutually exclusive, because their ejecta are not mixed at the microscopic level. Thus we evaluate the mass condensed into dust grains in SNII ejecta as
\begin{equation}
    M_{\rm C\, dust} = \delta_{\rm SNII, C} M_{\rm C \, ej}
\end{equation}
for carbon grains and
\begin{equation}
    M_{ \rm X\, dust} = N_{\rm sil} \mu_{\rm X} N_{\rm ato}^{\rm X}.
\end{equation}
for silicates, being ${\rm X} \in \text{[Mg,Fe,Si,O]}$. In above expressions, $M_{\rm X \, ej}$ is the mass of the ${\rm X}$ element ejected by SNII, and $N_{\rm sil}$ is computed as in Eq. \ref{eq:Nsil}, adopting $\delta_{\rm SNII, \, C}=\delta_{\rm SNII, \, sil}=0.1$.

\subsubsection{Shattering and coagulation}
\label{sec:dustshatcoa}

When the relative velocity of large grains, originated by ISM turbolence \citep[e.g.][]{Yan2004}, is high enough ($v \sim 1-10 \, \text{km\, s}^{-1}$), such particles may collide and fragment. This process, known as \textit{shattering} (e.g. \citealt{Hirashita2009}), originates small grains. In the context of the two-size approximation adopted here, shattering simply transfers mass from large to small grains. The corresponding timescale in our model is computed for each cold gas ring as (see \citealt{Granato2021}):
\begin{equation}
    \tau_{\rm sh }=
    \label{eq:shat}
    \begin{cases}
    \tau_{\rm sh,0} \left(\frac{0.01}{\rm DTG_{\rm L} } \right)\frac{\rm cm^{-3}}{n_{\rm gas}} \qquad \qquad \qquad \qquad n_{\rm gas} < 1 \, \text{cm}^{-3}, \\
    \tau_{\rm sh,0} \left(\frac{0.01}{\rm DTG_{\rm L}}\right) \frac{\rm cm^{-3}}{n_{\rm gas}} \left(\frac{n_{\rm gas}}{\rm cm^{-3}}\right)^{2/3} \qquad 1 \, \text{cm}^{-3}< n_{\rm gas} < 10^3 \, \text{cm}^{-3}, \\
    \end{cases}
\end{equation}
where $\tau_{\rm sh,0}= 5.41 \cdot 10^7 \, {\rm yr}$, DTG$_{\rm L}$ is the dust-to-gas ratio considering large grains only, and $n_{\rm gas}$ is the gas density. Clearly, in our SAM some assumption is needed to use a proper gas density in the above formula. In our fiducial model, we compute $n_{\rm gas}$ in each ring from:

\begin{equation}    
    \rho_{\rm gas} = \dfrac{\Sigma_{\rm cold \, gas}}{0.1 R_{\rm cold \, gas}},  
    \label{eq:rhoshat}
\end{equation}
where $\Sigma_{\rm cold \, gas}$ is the surface density of the cold gas in a ring, and $R_{\rm cold \, gas}$ the radius of the cold gas disc (i.e. we assume the height of the disc to be $0.1$ its radius). We derive $n_{\rm gas}$ from the above formula assuming $\mu = 1.2$. The reasons for the choice of this prescription instead of, for example, a fixed $n_{\rm gas}$ are briefly discussed in Appendix \ref{app:SLprofile}.

By converse, in the densest phases of the ISM, which are molecular clouds, small grains have relative velocities low enough to allow them to aggregate into large grains. This process, dubbed \textit{coagulation} (e.g. \citealt{Hirashita2009}), transfers mass from small to large grains, without affecting the total dust budget of the cold gas. For its implementation, we follow \cite{Aoyama2017} (see also \citealt{Granato2021}) and derive a timescale:
\begin{equation}
    \tau_{\rm coa} = \tau_{\rm co,0} \left(\frac{0.01}{\rm DTG_S}\right) \left(\frac{0.1\cdot \text{km s}^{-1}}{v_{\rm coa}}\right) \frac{1}{f_{\rm H_2}}. 
\end{equation}
In the above expression we set $\tau_{\rm coa,0}=2.71 \cdot 10^5 \, \text{yr}$, $v_{\rm coa}=0.2 \, \text{km s}^{-1}$, DTG$_{\rm S}$ is the dust-to-gas ratio considering only small grains, and $f_{\rm H_2}$ is the molecular gas fraction of the ring.

\subsubsection{Accretion}
\label{sec:dustgrowth}
In the coldest and densest phases of the ISM, gas phase metals can stick on the surface of dust grains, accreting their mass (e.g. \citealt{Dwek1998}). This is the \textit{grain growth} or \textit{accretion} process and is expected to be efficient in molecular clouds. In this work, we model the accretion time-scale for each element ${\rm X}$ following \cite{Hirashita2011} (see also \citealt{Granato2021}):
\begin{equation}
\label{eq:acc}
    \tau_{{\rm acc}, {\rm X}} = \frac{a f_{\rm X} s \mu_{\rm X} }{3 n Z_{\rm X} \bar{\mu} S} \left(\frac{2 \pi}{m_{\rm X} k T}\right)^{0.5} \frac{1}{f_{\rm H_2}}.
\end{equation}

In the above expression, $S=0.3$ is the sticking efficiency, $T=50\, \rm{K}$ and $n=10^3 \, \text{cm}^{-3}$ the temperature and the density assumed for molecular clouds, $\mu_{\rm X}$ and $m_{\rm X}$ the atomic weight and mass of the element, $\bar{\mu}$ the mean molecular weight, $f_{\rm X}$ the mass fraction of the element in the grain, and $Z_{\rm X}$ the gas phase mass fraction of the element. The material density $s$ is assumed to be $s=2.2 \, \text{g cm}^{-3}$ and $s=3.3 \, \text{g cm}^{-3}$ for carbonaceous and silicate grains. $f_{\rm H_2}$ is the molecular gas fraction of the ring, and $a$ is the grain radius. Following \cite{Hirashita2015}, we assume this surface process works only for small grains, since the surface-mass ratio is larger for them. We use the representative radius $a=0.005 \, \mu\text{m}$ for small grains in the previous equation \cite[see][]{Granato2021}.

The accretion timescale is evaluated at each timestep, for each element, and in each ring. For carbon grains, we adopt the timescale obtained with Eq. \ref{eq:acc}. For the elements of silicate grains (O, Si, Mg, and Fe) we adopt the accretion timescale of the element which maximizes it; this is needed to preserve the olivine-like chemical composition adopted in this work.

\subsubsection{Destruction in SNe shocks}
\label{sec:dustSNshock}
Dust grains may be efficiently eroded in shocks derived from SNae explosions, primarily by kinetic sputtering.
Our model of dust destruction in SNae shocks is inspired by \cite{Asano2013} and acts on both small and large grains. For each cold gas ring, we evaluate the timescale of the process as:
\begin{equation}
    \tau_{\rm des, SN} = \frac{M_{\rm cold}}{\epsilon_{\rm SN} M_{\rm swept} R_{\rm SN}},
\end{equation}
where $M_{\rm cold}$ is the mass of the gas ring, $M_{\rm swept}$ is the gas mass swept by a SN event, $R_{\rm SN}$ is the SN rate and $\epsilon_{\rm SN}$ is a parameter quantifying the dust destruction efficiency. Here we follow \cite{Asano2013} (see also \citealt{Yamasawa2011}) in evaluating the  swept mass:
\begin{equation}
    M_{\rm swept } = 1535 \left( \frac{Z}{Z_\odot} + 0.039\right)^{-0.289} \, M_\odot,
\end{equation}
being $Z$ the gas metallicity of the ring. The SN rate is computed self-consistently by the SAM, and here we do not differentiate among SNII and SNIa, so that $R_{\rm SN} = R_{\rm SNII}+R_{\rm SNIa}$\footnote{We only consider SNe from disc stars.}. Finally, we assume a destruction efficiency $\epsilon_{\rm SN}=0.1$ (\citealt{McKee1989}).

\subsubsection{Sputtering}
\label{sec:dustspu}
In the hot gas, grains may be eroded by \textit{thermal sputtering}, namely collisions between grains and energetic particles. We model it through a timescale (\citealt{Tsai1995}; \citealt{Popping2017}; \citealt{Granato2021}):
\begin{equation}
\label{eq:tauspu}
    \tau_{\rm spu} = \tau_{\rm spu,0} \frac{a_{0.1 \, \rm{\mu m}}} {\rho^{\rm gas, hot}_{10^{-27} \, \text{g cm}^{-3}}} \left[\left(\frac{T_{\rm spu,0}}{{\rm min}(T_{\rm gas,hot}, 3 \cdot 10^7 \, \text{K})} \right)^\omega +1 \right],
\end{equation}

where $\tau_{\rm spu,0}= 0.17/3 \, \text{Gyr}$, $T_{\rm spu,0}= 2 \cdot 10^6 \, \text{K}$, $\omega = 2.5$, $a_{0.1 \, \rm{\mu m}}$ is the grain radius in $0.1 \, \rm{\mu m}$ units, $\rho^{\rm gas, hot}_{10^{-27} \, \text{g cm}^{-3}}$ the hot gas density in $10^{-27} \, \text{g cm}^{-3}$ units, and $T_{\rm gas, hot}$ the hot gas temperature, which we assume to be the virial temperature. The representative grain radii assumed for large and small grains are, respectively, $a_{\rm L}=0.05\, \mu{\rm m}$ and $a_{\rm S}=0.005\, \mu{\rm m}$.

\rev{The above empirical fitting formula captures, besides the increase of ion-grain collisions with plasma temperature and density, the $T$-dependence 
of sputtering yield resulting from theoretical computation \citep[e.g.][]{Tielens1994,Nozawa2006}\footnote{See also Appendix A of \cite{McKinnon2017}.}.
At low $T$ the erosion rate rapidly increases with $T$, then it flattens above $\sim T_{\rm spu,0}$.
Moreover, since sputtering is a surface process, its timescale is proportional to grain radius.} 

\rev{In Eq.\ \ref{eq:tauspu} we simply adopt for the hot gas density the average value obtained assuming that it fills the DM halo up to the virial radius and for the temperature the virial temperature $T_{\rm vir}=\dfrac{\mu m_{\rm p}}{2 k} V^2_{\rm vir}$:}
\begin{equation}
    \rho^{\rm gas, hot} = \frac{M_{\rm gas, hot}}{4\pi R^3_{\rm vir}/3}
\quad {\rm and} \quad
    T_{\rm gas, hot} = 35.9 \cdot \left( \frac{V_{\rm vir}}{\rm km/s}\right)^2 \, [{\rm K}],  
\end{equation}
where $R_{\rm vir}$ and $V_{\rm vir}$ are the virial radius and virial velocity of the halo for central galaxies, or these quantities at infall for satellites. 
The same prescription is adopted for dust in the ejected reservoir.
In our fiducial model, we adopt a $\tau_{\rm spu,0}$ larger by a factor $10$, i.e. we reduce the sputtering efficiency to provide a better match of the cosmic abundance of CGM dust (see discussion in Sec. \ref{sec:Omega_hot}). This choice only influences dust in the hot gas and ejected reservoir. In other words, the impact of sputtering on the cold phase dust is negligible.

\subsection{Disc Instabilities}
\label{sec:DImodel}
We update the criterion for disc instabilities originally adopted in \textsc{L-Galaxies 2020}, which is based on \cite*{Efstathiou1982}. This prescription aims at determining when a disc is unstable to bar formation due to its self-gravity; this instability then re-distributes the disc material, eventually funnelling material towards a bulge-like structure (e.g. \citealt{Debattista2006}). As a result of its attractive simplicity, this criterion has been widely adopted, often with minor modifications, by the SAM community (e.g. \citealt{Lucia2011}; \citealt{Lacey2016}; \citealt{PLagos2018}), though not without criticism (\citealt{Athanassoula2008}; \citealt{Devergne2020}; \citealt{Romeo2022}). This secular process is fundamental to produce enough intermediate mass spheroids (e.g. \citealt{Husko2022}), and it may also play a relevant role in BH growth at high redshift (\citealt{Bournaud2011}).

Our implementation is inspired to\footnote{Differently from \cite{Irodotou2019}, we do not distinguish between classical bulges (formed during mergers) and pseudo-bulges (originated from disc instabilities). Another crucial difference is the division into rings of discs in the SAM we adopt. Here we determine if a disc is unstable according to its global properties, but a ring-by-ring evaluation of the instability would also be an interesting experiment (see e.g. \citealt{Stevens2016}).} 
\cite{Irodotou2019} and extend the instability criterion in order to take into account the contribution of the gaseous disc to the stability of the whole disc (gas+stars; only stars are considered in \citealt{Henriques2020}).
To do this, we evaluate the stability of the galactic disc by mean of the parameter $\epsilon_{\rm tot}$:
\begin{equation}
    M_{\rm disc, tot} \epsilon_{\rm tot} = M_{\rm disc,stars}\epsilon_{\rm stars} + M_{\rm disc,gas}\epsilon_{\rm gas},
\end{equation}
where
\begin{equation}
    \epsilon_{i} = c_i \left( \frac{GM_{\rm disc, i}}{V^2_c R_{\rm disc, i}}\right)^{0.5},
\end{equation}
with $i=\{\rm stars, gas \}$. 
In the above equation, $V_c$ is the circular velocity of the host halo, $M_{i}$ is the mass of the $i$ component of the disc, $R_{i}$ the scale length, and $c_{\rm stars}$, $c_{\rm gas}$ are parameters of order of 1 (\citealt{Efstathiou1982}; \citealt{Christodoulou1995}). In our fiducial model we set $c_{\rm stars}=c_{\rm gas}=1$.\\

If $\epsilon_{\rm tot}>1$ we consider the disc to be unstable. We then allow the gas to form stars and to accrete the central SMBH, and move stars from disc to bulge, until stability is restored.
This is done by iterating over rings, from the innermost to the outermost one, and for each of them:

\begin{itemize}
\item if gas $M_{\rm gas}^{\rm ring}$ is present in the ring, we allow some of it to form disc stars and to accrete the central SMBH. In a single timestep $dt$, we assume that the mass available for the former processes is $M_{\rm gas, unst}^{\rm ring} = (dt/\tau) \cdot  M_{\rm gas}^{\rm ring}$, where $\tau$ is proportional to the free fall timescale of each ring, i.e. 
    \begin{equation}
        \tau = \sqrt{\frac{1}{G\rho}},
    \end{equation}
where $\rho$ is evaluated as in Eq. \ref{eq:rhoshat}.\\
A fraction $f_{\rm BH, unst}$ of the mass $M_{\rm gas, unst}$  accretes onto the central SMBH. This fraction is evaluated following almost the same prescription adopted by the SAM when dealing with BH growth in galaxy mergers\footnote{Eq. S27 of the supplementary material of \citealt{Henriques2020}, setting $M_{\rm sat}/M_{\rm cen } = 1$.}, namely:
\begin{equation}
    \label{eq:BHaccinsta}
    f_{\rm BH, unst} = \frac{f_{\rm BH}}{1 + \left(\dfrac{V_{\rm BH,\, DI}}{V_{\rm vir}}\right)^2},
\end{equation}

\rev{where $f_{\rm BH}$ controls the accretion efficiency and the $(V_{\rm BH,\, DI}/V_{\rm vir})^2$ term introduces a dependence on the binding energy of the system since we expect that less gas concentrates in the center of low mass systems.}

\rev{This prescription was introduced initially by \cite{Kauffmann2000} to model the gas accretion of SMBH during major mergers. Such \textit{phenomenological} recipe has been then widely adopted by the SAM community (e.g. \citealt{Croton2016}; \citealt{PLagos2018}; \citealt{Izquierdo-Villalba2020}) in both galaxy mergers and disc instabilities when gas is funnelled towards the centre of the galaxy and gives origin to starburst episodes. We thus adopt the same approach. In particular, in our fiducial model, during disc instabilities we assume $f_{\rm BH}=0.066$ as in mergers, while we adopt a $V_{\rm BH, \, DI}$\rev{, that is the virial velocity at which accretion saturates,} larger by a factor of $6$ than the one adopted during mergers ($700 \, \rm{km \, s^{-1}}$). This increase is required to avoid excessive growth of SMBHs and ensuing quenching of intermediate mass galaxies. The role of the $V_{\rm BH, \, DI}$ parameter is discussed in Appendix \ref{app:BHacc}. Moreover, there we discuss tests simply adopting  a constant $f_{\rm BH, unst}$. We found that $f_{\rm BH, unst}=10^{-4}$ approxinately reproduces the most important results of our work. Indeed, some other SAMs (e.g. \citealt{Lacey2016}) assume that a fixed fraction of the unstable gas accretes on the SMBH.}

Once the fraction of the unstable mass accreting the BH has been determined, the residual fraction $1 - f_{\rm BH, unst}$ form stars. \\

\item If a stellar mass $M_{\rm stars} ^{\rm ring}$ is in the disc ring, we move it to the bulge. In a single timestep $dt$, we perform a sub-iteration of $N_{\rm steps}$, removing at each sub-step a mass  $M_{\rm stars} ^{\rm ring}/N_{\rm steps}$ and re-evaluating the stability of the disc each time. This procedure avoids removing the whole stellar mass from a single ring when just a fraction of it would be sufficient to restore stability. We adopt $N_{\rm steps}=20$, and we verified any value $N_{\rm steps}\gtrsim10$ yields similar results.
\end{itemize}

We perform the steps above for each ring until the stability of the whole disc is restored. We update bulge sizes and bulge masses simultaneously. Since bulges are not the main topic of this paper, we refer the interested reader to Appendix \ref{app:sizes}, where the computation of bulge sizes is detailed and few results are shown.\\

\section{Results}
\label{sec:results}
In the following sections, we present and discuss the main results of our simulations concerning the general properties of the galaxy population (Sec. \ref{sec:sanity} and \ref{sec:morphoSF}), their dust content (Sec. \ref{sec:dust_scaling}) and the cosmic dust density (Sec. \ref{sec:dust_cosmic}). We will mainly focus on two runs, dubbed \textit{FID} (our fiducial model) and \textit{oldInsta}. Both embed our dust modelling introduced in Sec. \ref{sec:dustmodel}. However, only the former relies on the prescriptions for disc instability detailed in Sec. \ref{sec:DImodel} (in \textit{oldInsta} the same prescriptions of \citealt{Henriques2020} are adopted). \\

\subsection{General properties} 
\label{sec:sanity}
\begin{figure*}
    \centering
    \begin{subfigure}[b]{0.3\textwidth}
        \includegraphics[width=\textwidth]{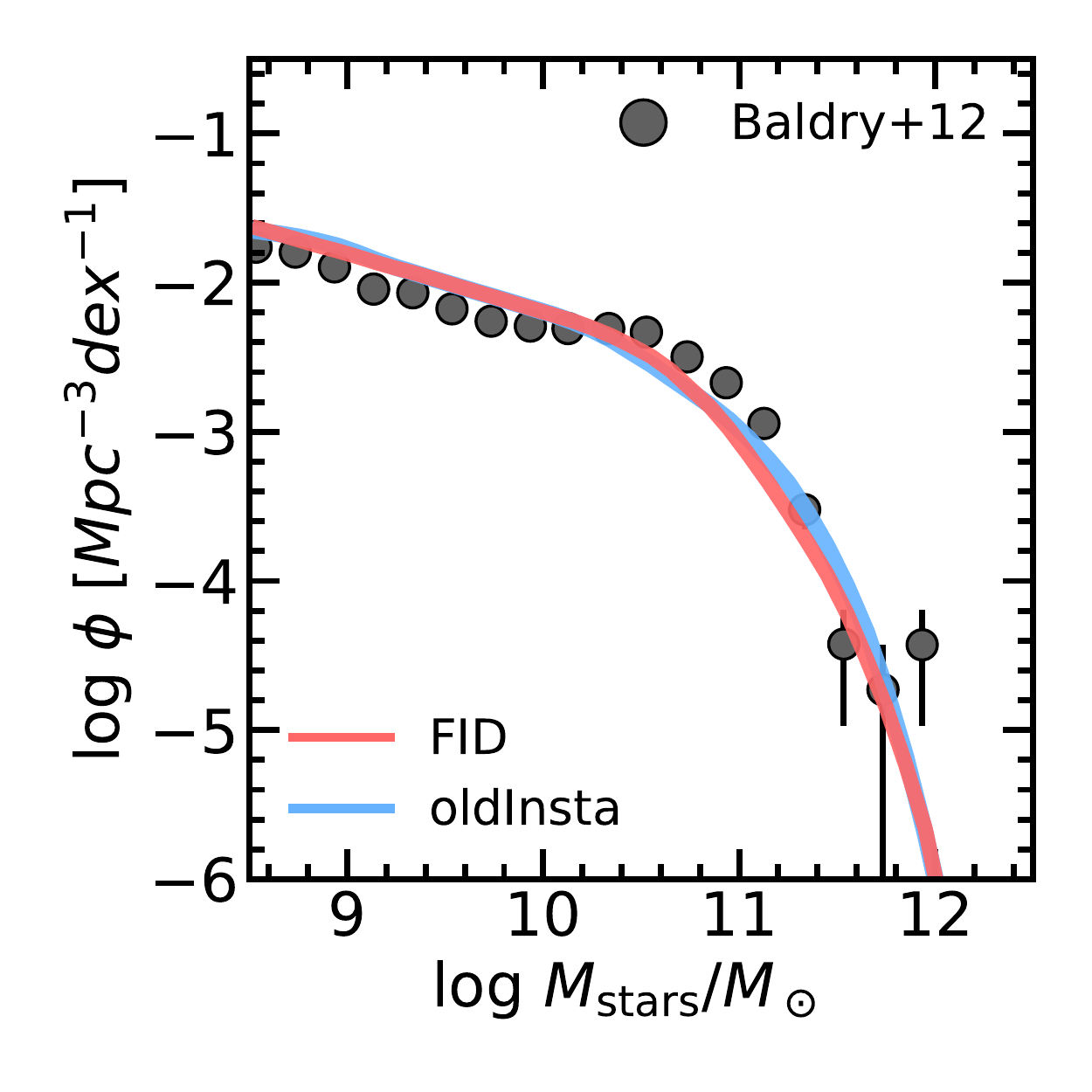}
        \caption{}
        \label{fig:SMF}
    \end{subfigure}
    \begin{subfigure}[b]{0.3\textwidth}
        \includegraphics[width=\textwidth]{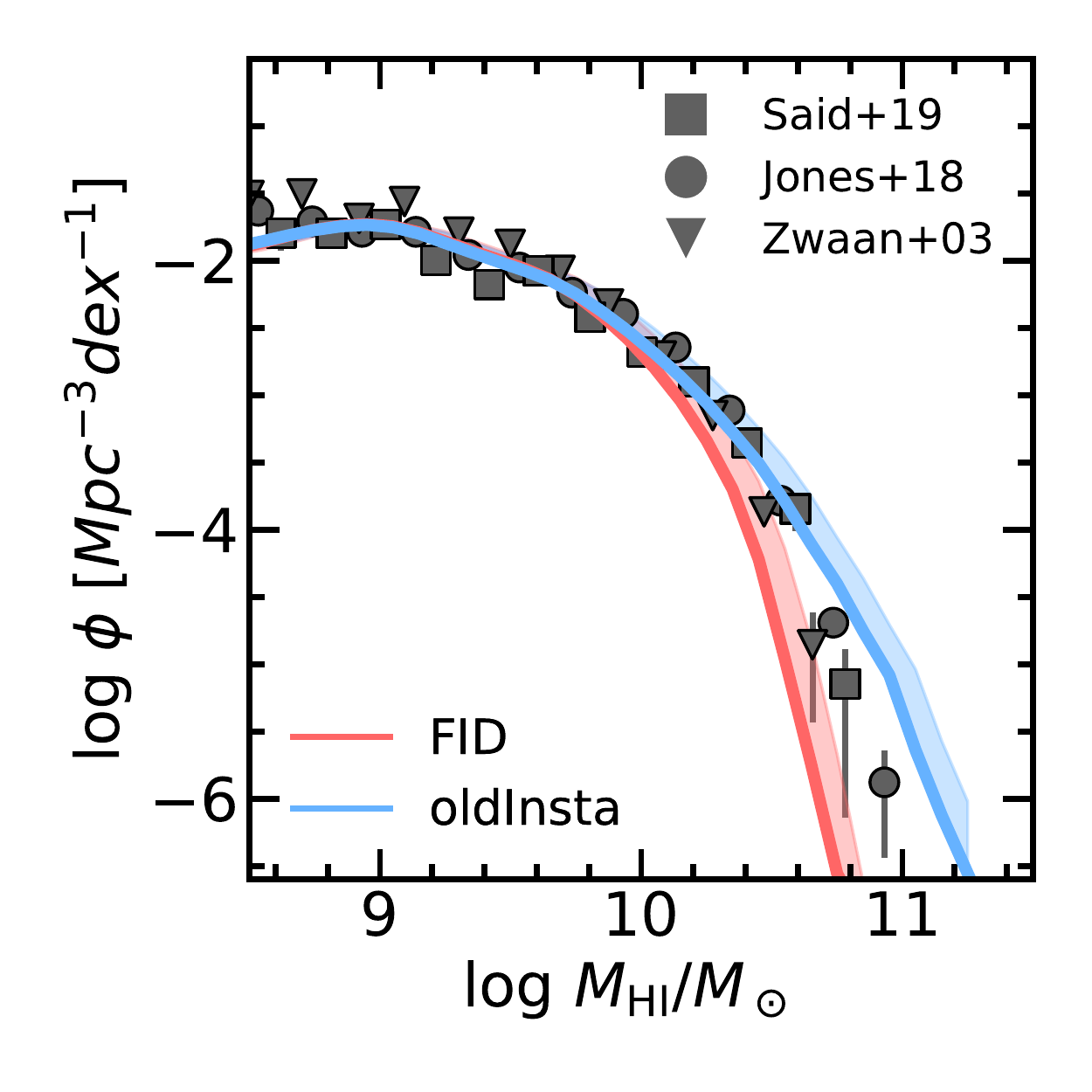}
        \caption{}
        \label{fig:HIMF}
    \end{subfigure}
    \begin{subfigure}[b]{0.3\textwidth}
        \includegraphics[width=\textwidth]{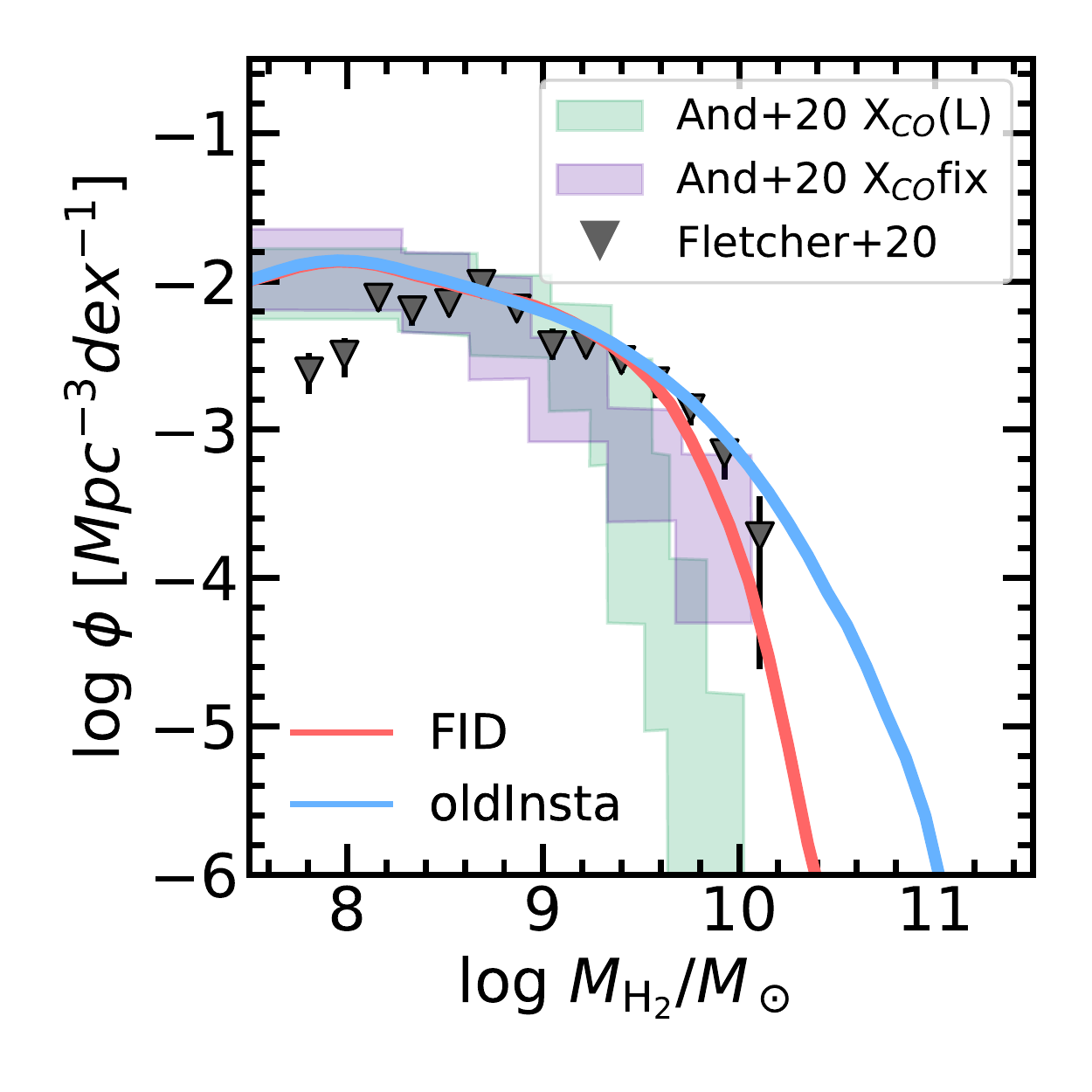}
        \caption{}
        \label{fig:H2MF}
    \end{subfigure}
    \caption{Mass functions at $z=0.0$ of our fiducial model (\textit{FID}; red) and the model without updated treatment of disc instabilities (\textit{oldInsta}; blue). \textit{Left panel:} SMF compared with observations by \citet{Baldry2012}. \textit{Central panel:} neutral hydrogen mass function (HIMF) compared with data from \citet{Zwaan2003}, \citet{Jones2018}, and \citet{Said2019}. Here we show the dispersion obtained assuming the neutral fraction to be $1/1.3$ (\citealt{Fu2010}) or $1$. \textit{Right panel:} molecular hydrogen mass function compared with observations by \citet{Fletcher2020} and \citet{Andreani2020}. As for the latter work, we show their results obtained assuming either a constant (purple) and luminosity-dependent (green) CO conversion factor.}
    \label{fig:MFmain}
\end{figure*}

\begin{figure}
    \centering

    \includegraphics[width=0.9\linewidth]{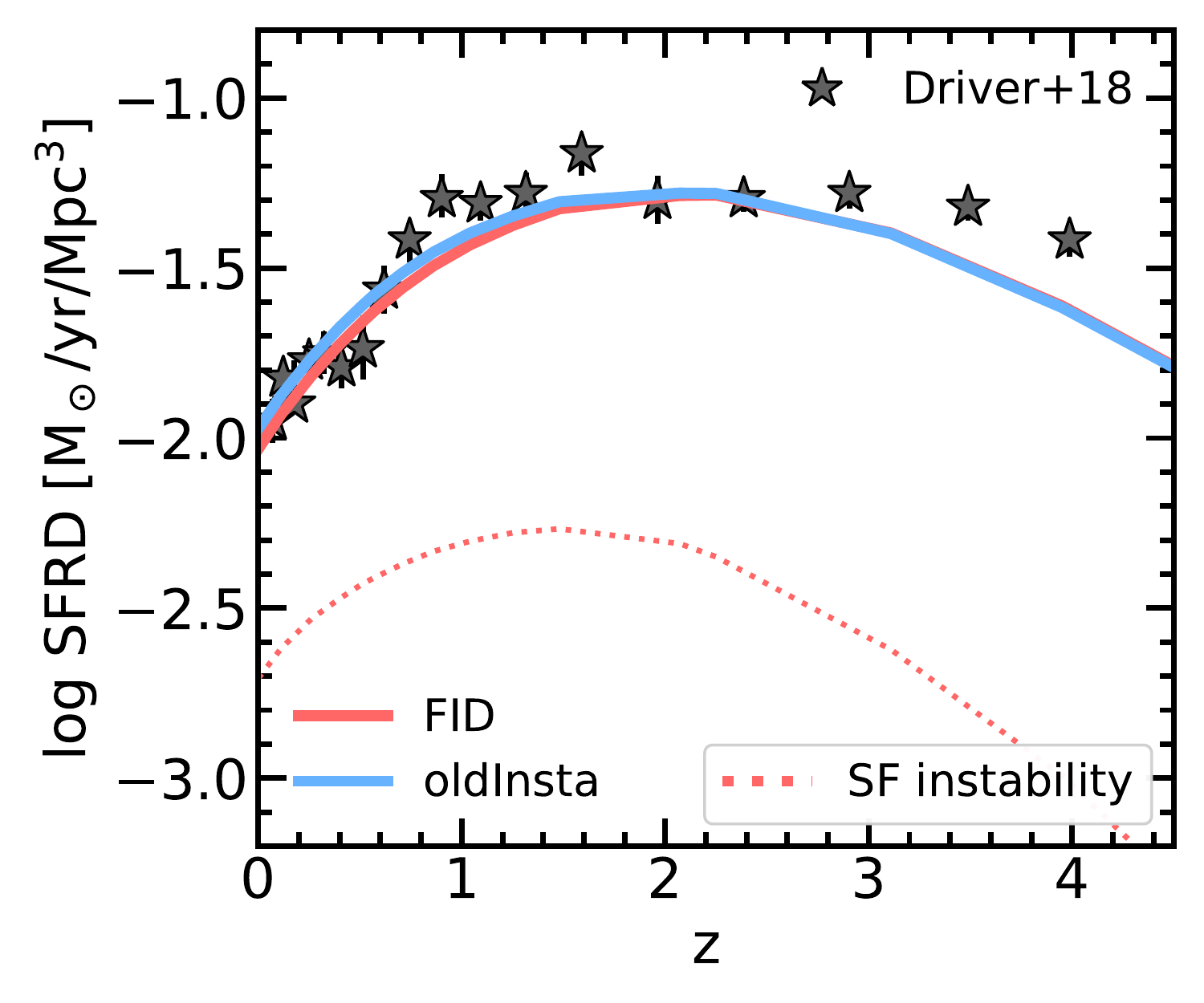}

    \caption{Cosmic evolution of the Star Formation Rate Density in our fiducial model (\textit{FID}; red) and in the model without the new implementation of disc instabilities (\textit{oldInsta}; blue). \rev{As for our FID model, we also show the star formation triggered by disc instability episodes (SF instability; dotted line).} Data from \citet{Driver2018} are also shown for comparison.}
    \label{fig:SFRD}
    
\end{figure}

As a first benchmark for our model, in this section we show its results for some fundamental quantities: the Stellar, HI and H$_2$ Mass Function at $z=0.0$ (respectively SMF, HIMF and H$_2$MF) and the cosmic Star Formation Rate Density (SFRD). In our fiducial model, the first two quantities (Figg. \ref{fig:SMF} and \ref{fig:HIMF}) slightly differ from what is obtained with the original instability criterion of \cite{Henriques2020} (\textit{oldInsta}). While the fiducial SMF is still in line with data, the HIMF underpredicts the observations at the high mass\footnote{This issue holds regardless of the uncertainty linked to the neutral gas fraction of the cold phase of a galaxy, here assumed to be $f_{\rm neu} = 1/1.3$ (\citealt{Fu2010}). Also, in the extreme case when $f_{\rm neu}=1$ the HIMF appears to be under-abundant at high masses.}. On the contrary, both the H$_2$MF (Fig. \ref{fig:H2MF}) and the cosmic SFRD (Fig. \ref{fig:SFRD}) are in good agreement with observations when our fiducial model is adopted. In particular, the former quantity is extremely improved when compared to the recent determinations of \cite{Andreani2020} and \cite{Fletcher2020}: the \textit{oldInsta} model has too much H$_2$ in the most massive galaxies, as already discussed in \cite{Henriques2020}. 

We anticipate the reason for the differences outlined above: our model allows for the SMBHs growth during disc instabilities, resulting in an earlier growth of SMBHs\footnote{Bulge and BH masses in our model remain consistent with observations (Appendix \ref{app:sanity2}).}, and then a faster quenching by cooling suppression due to the radio mode feedback. This results in lower gas contents (and then H$_2$ and star formation) when disc instabilities occur.

In conclusion, our fiducial model produces a reasonable $z=0.0$ galaxy population and a  cosmic SFRD in good keeping with the data. 
This holds despite some modifications we introduced on the physics implemented in the SAM, without parameter re-tuning.


\subsection{Galaxy morphology and star formation}
\label{sec:morphoSF}

\begin{figure*}
    \centering
    \begin{subfigure}[t]{0.4\textwidth}
        \centering
        \includegraphics[width=\linewidth]{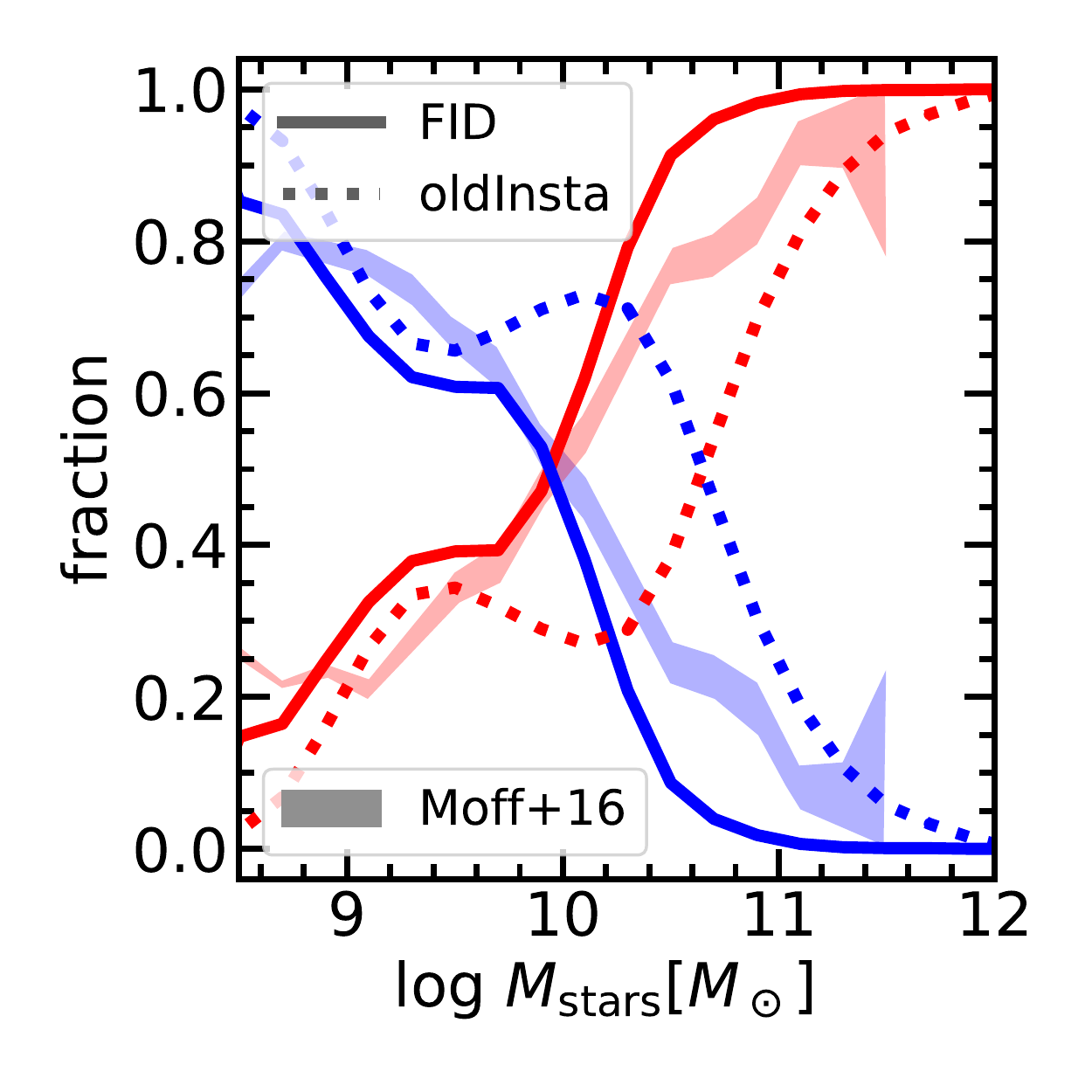}
        \caption{}
        \label{fig:morphoMstar}
    \end{subfigure}
    \begin{subfigure}[t]{0.48\textwidth}
        \centering
        \includegraphics[width=\linewidth]{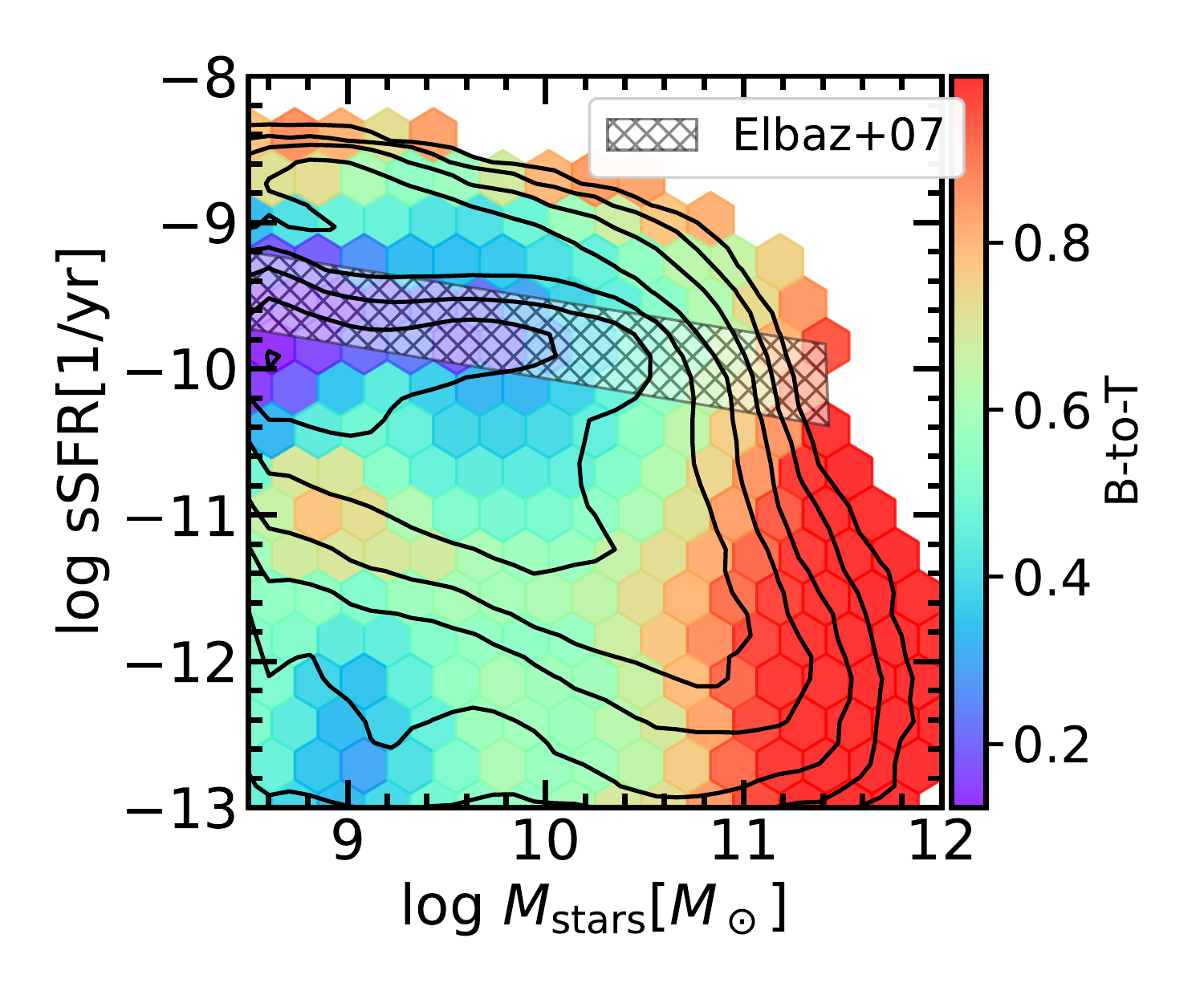} 
        \caption{}
        \label{fig:galMS}
    
    \end{subfigure}

    \caption{\textit{Left panel:} fraction of disc-dominated galaxies (B-to-T $ \leq 0.3$, blue) and spheroid-dominated galaxies (B-to-T $>0.3$, red) for our fiducial model (\textit{FID}; solid line) and the model without the updated disc instability model (\textit{oldInsta}; dotted line). We compare with observations of \citet{moffett2016} (coloured regions).
    \textit{Right panel:} Specific SFR as a function of stellar mass at $z=0.0$ for our fiducial model. The bins of the 2D histogram are coloured according to the mean B-to-T, while log-spaced density contours are shown as black lines to give a qualitative idea of the distribution of galaxies in this plane. We overplot the \citet{Elbaz2007} SF main sequence as a gray hatched region.} 

    \label{fig:morpho}
    
\end{figure*}

In this section, we first inspect the morphology of the model galaxies, quantified using the Bulge-to-Total stellar mass ratio (B-to-T). We show in Fig. \ref{fig:morpho} the relative fraction of disc-dominated (B-to-T $<0.3$) and spheroid-dominated galaxies (B-to-T $>0.3$) as a function of stellar mass for our fiducial model, as well as for the model without the new implementation of disc instabilities (\textit{oldInsta}). In the same figure, we show the results of the Galaxy And Mass Assembly survey (GAMA; \citealt{moffett2016}). In their local sample of observed galaxies, these authors study the contribution of different morphological types to the total galaxy stellar mass function. They found that the transition point marking the dominance of spheroidal galaxies (E, S0-Sa, LBS\footnote{We include Little Blue Spheroids (LBS) in spheroidal-dominated galaxies, differently from what is done by \cite{moffett2016}. LBS are often associated with pseudo-bulges (Sérsic index $n\leq 2$). Since we do not explicitly discriminate between classical and pseudo-bulges, we include LBS in spheroid-dominated galaxies. However, we point out that their contribution is only relevant at $M_{\rm stars}\lesssim 10^{9.5} \, M_\odot$.}) types over disc-dominated galaxies (Sd-Irr, Sab-Scd types) occurs at $M_{\rm stars} \simeq 10^{10} \, M_\odot$. This finding is in excellent agreement with our model predictions. The updated disc instability recipe is required to produce a reasonable number of bulge-dominated systems since the \textit{oldInsta} model predicts a too large value of the transition mass. Disc instabilities turn out to be fundamental for bulge formation in the $10^{10}-10^{11} \, M_\odot$ stellar mass range, where bulge growth by mergers is insufficient (see the detailed discussion in \citealt{Irodotou2019}, in the framework of \textsc{L-Galaxies 2015}, or in \citealt{Husko2022}) and near the mass range ($3 \cdot 10^9 < M_*/M_\odot < 3 \cdot 10^{10}$) in which according to observations pseudobulges dominate \citep{Fisher2011}.\\

Once we verified that our model reproduces the number of spheroidal galaxies observed in the local Universe, we now analyze their star formation properties. The link between galactic morphology and galaxy properties has been observed and studied for a long time (e.g. \citealt{Roberts1994}), with early type galaxies being typically redder and less star forming than late-type ones (see e.g. the recent work by \citealt{Dimauro2022}). Although a solid theoretical explanation for this is still missing, many SAMs (including the present one) produce passive, elliptical galaxies as a result of mergers, which trigger bulge formation and BH growth. The subsequent BH radio mode feedback prevents the cooling of the hot gas and consequently stops the star formation (e.g. \citealt{Croton2006}). In our specific case, we introduce a new BH-growth channel, namely gas accretion during disc instabilities, and then we indirectly modify the original quenching scheme. 

It is thus interesting to inspect the star formation properties of our simulated galaxies. First, in Fig. \ref{fig:galMS} we show the $z=0$ distribution of our galaxies in the specific SFR (sSFR$=\rm{SFR}/M_{\rm stars}$) - $M_{\rm stars}$ plane, coloured according to the mean B-to-T in each bin. We also overplot the fit to the observed SF main sequence proposed by \cite{Elbaz2007}. Our simulated galaxies follow this relation up to log $M_{\rm stars}/M_\odot \lesssim 10^{10.5}$. Galaxies overlapping to the observed main sequence are principally disc-like (B-to-T $\lesssim 0.4$). With increasing stellar mass, the passive population becomes dominant. This population typically features an elliptical structure (B-to-T $\gtrsim 0.7$). Our model thus produces a qualitative correlation between the B-to-T and the position in the sSFR - $M_{\rm stars}$ plane, in line with observations (e.g. \citealt{Dimauro2022}). However, a detailed analysis of the interplay between morphology and quenching goes beyond the scope of the present work. We refer to \cite{Koutsouridou2022} for an investigation of this topic in a SAM.

\begin{figure*}
    \centering

    \includegraphics[width=1.\linewidth]{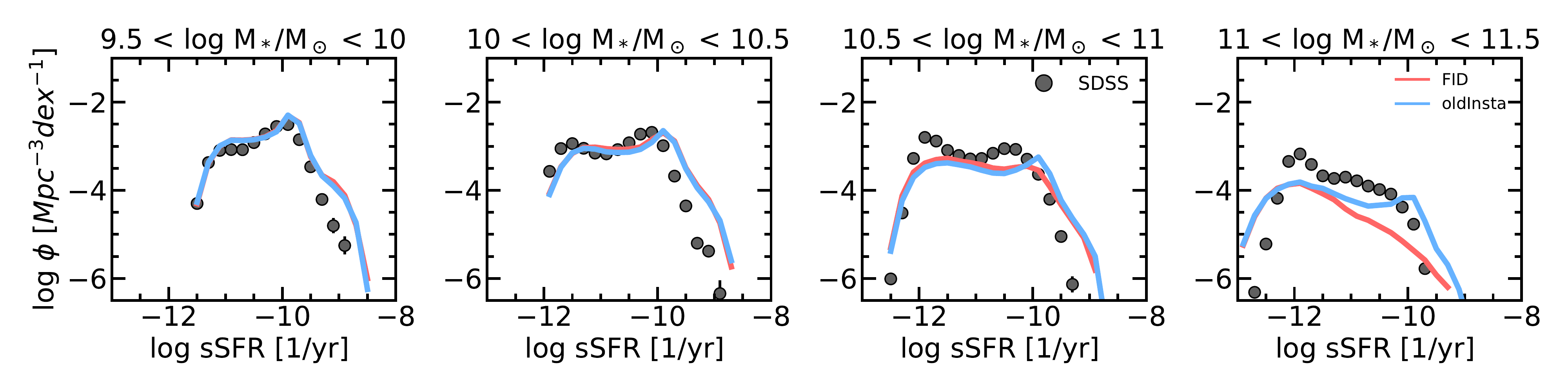}

    \caption{Specific SFR distribution for our $z=0.0$ sample of galaxies with log SFR$>10^{-1.5} \, M_\odot/{\rm yr}$ in stellar mass bins of width $\Delta {\rm log} M_*/M_\odot = 0.5$. Our model (\textit{FID}; red) is compared with the model without the updated implementation of disc instabilities (\textit{oldInsta}; blue). Data from SDSS-DR7 (\citealt{Katsianis2020}) are shown in each panel as filled circles.}
    \label{fig:histo-sSFR}
    
\end{figure*}

A deeper analysis of the impact of the new disc instability prescription on the star formation properties of our simulated galaxies is in Fig. \ref{fig:histo-sSFR}. There, we show the specific SFR  distribution in different stellar mass bins for the \textit{FID} and \textit{oldInsta} model, compared with SDSS-DR7 data (\citealt{Katsianis2020}). The two models perform similarly for log $M_{\rm stars}/M_\odot < 10.5 $, while at larger stellar masses the \textit{FID} model is characterized by generally lower sSFRs and less bimodality. Our fiducial model reproduces the shape of the observed distribution in the most massive stellar mass bin, although with a lower normalization. On the other hand, in the same mass bin the \textit{oldInsta} model predicts too much star forming objects, likely because of their excessive molecular gas content (see Fig. \ref{fig:H2MF}). \\

Star formation differences between these two models are due to the enhanced SMBH growth during disc instabilities\footnote{A few results for different choices of $V_{\rm BH, \, DI}$ in Eq. \ref{eq:BHaccinsta} are shown in Appendix \ref{app:BHacc}.}: BHs in unstable galaxies grow faster and prevent the cooling with their radio-feedback, leading to more rapid quenching of the galaxy.
This mechanism prevents the formation of massive, gas rich, star forming (and dust rich, see Sec. \ref{sec:DMF}) galaxies at $z=0.0$. However, the discussion above suggests that the current BH-driven quenching should be further revised to match SDSS data for ${\rm log }M_*/M_\odot > 11$, where our normalization is too low.


\subsection{Dust content of galaxies}
\label{sec:dust_scaling}
To validate our dust implementation within the SAM, we now discuss some model predictions concerning dust at various redshifts. In this Section, we only consider galactic dust, which corresponds to dust in the cold gas of galaxies. In this work, our primary focus concerns global galaxy properties, and thus we limit to these quantities, postponing to a future work any detailed analysis of spatially resolved dust properties. 

\subsubsection{Dust Mass Function}
\label{sec:DMF}

\begin{figure*}
    \centering
    \begin{subfigure}[b]{0.3\textwidth}
        \includegraphics[width=\textwidth]{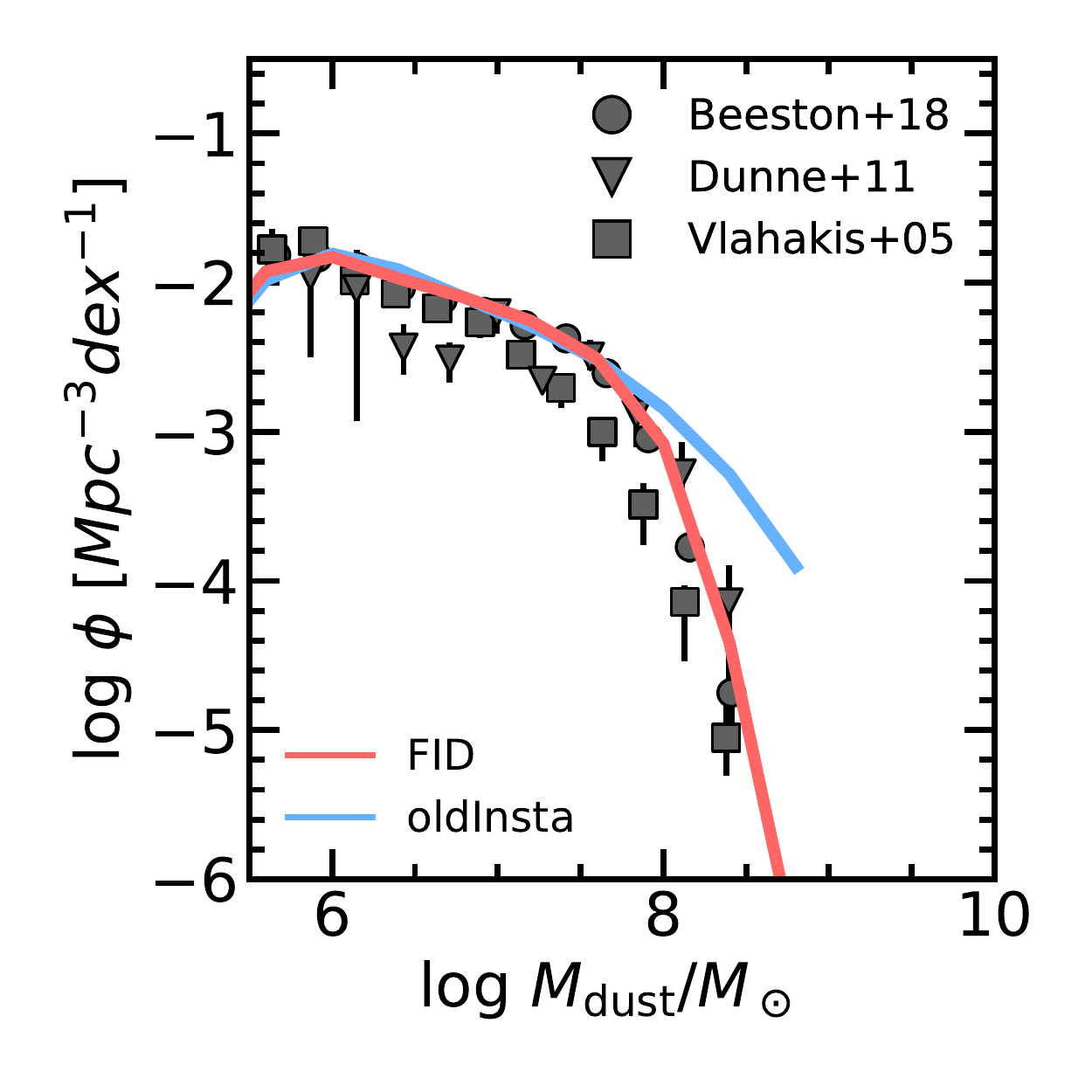}
        \caption{$z=0.0$}
        
    \end{subfigure}
    \begin{subfigure}[b]{0.3\textwidth}
        \includegraphics[width=\textwidth]{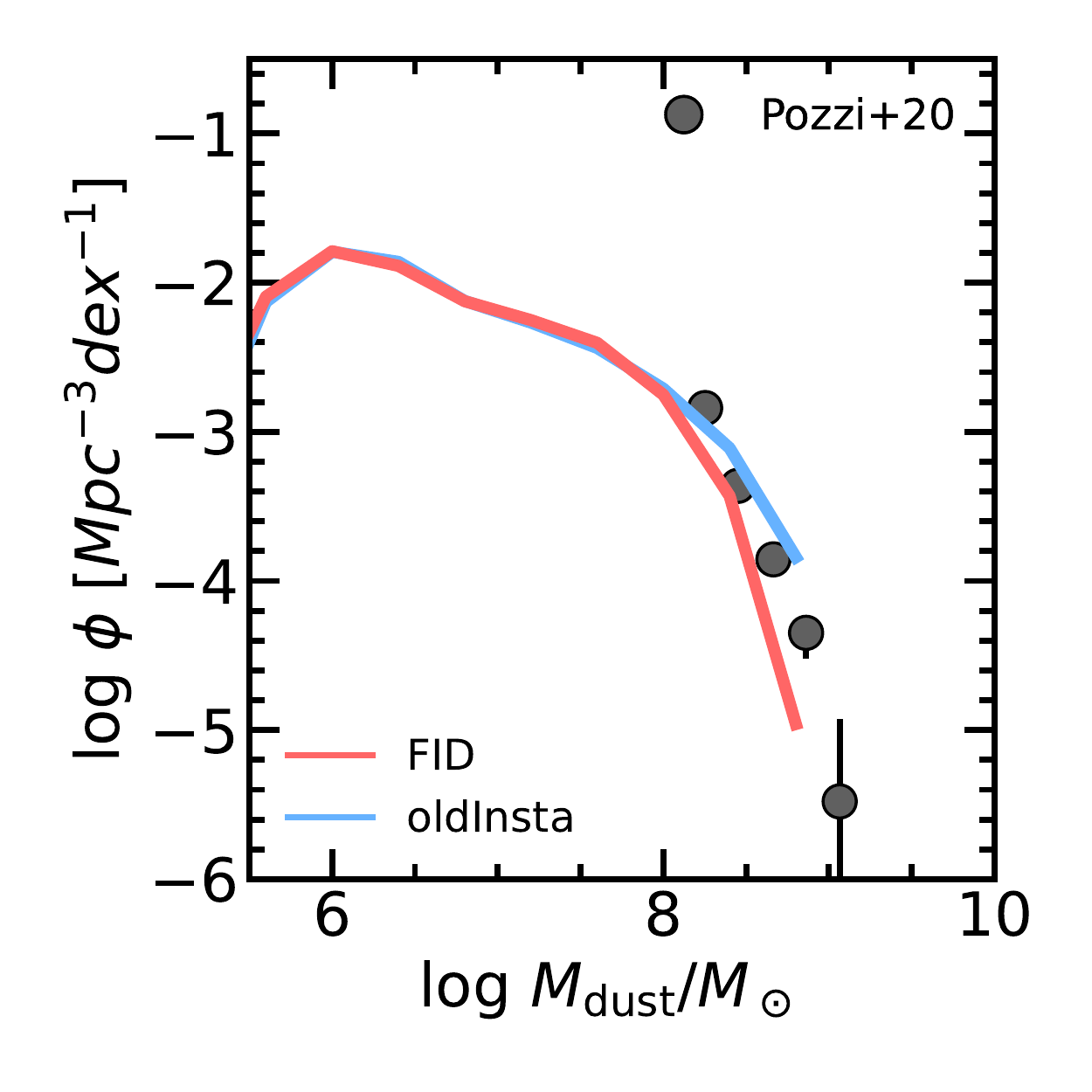}
        \caption{$z=1.0$}
       
    \end{subfigure}
    \begin{subfigure}[b]{0.3\textwidth}
        \includegraphics[width=\textwidth]{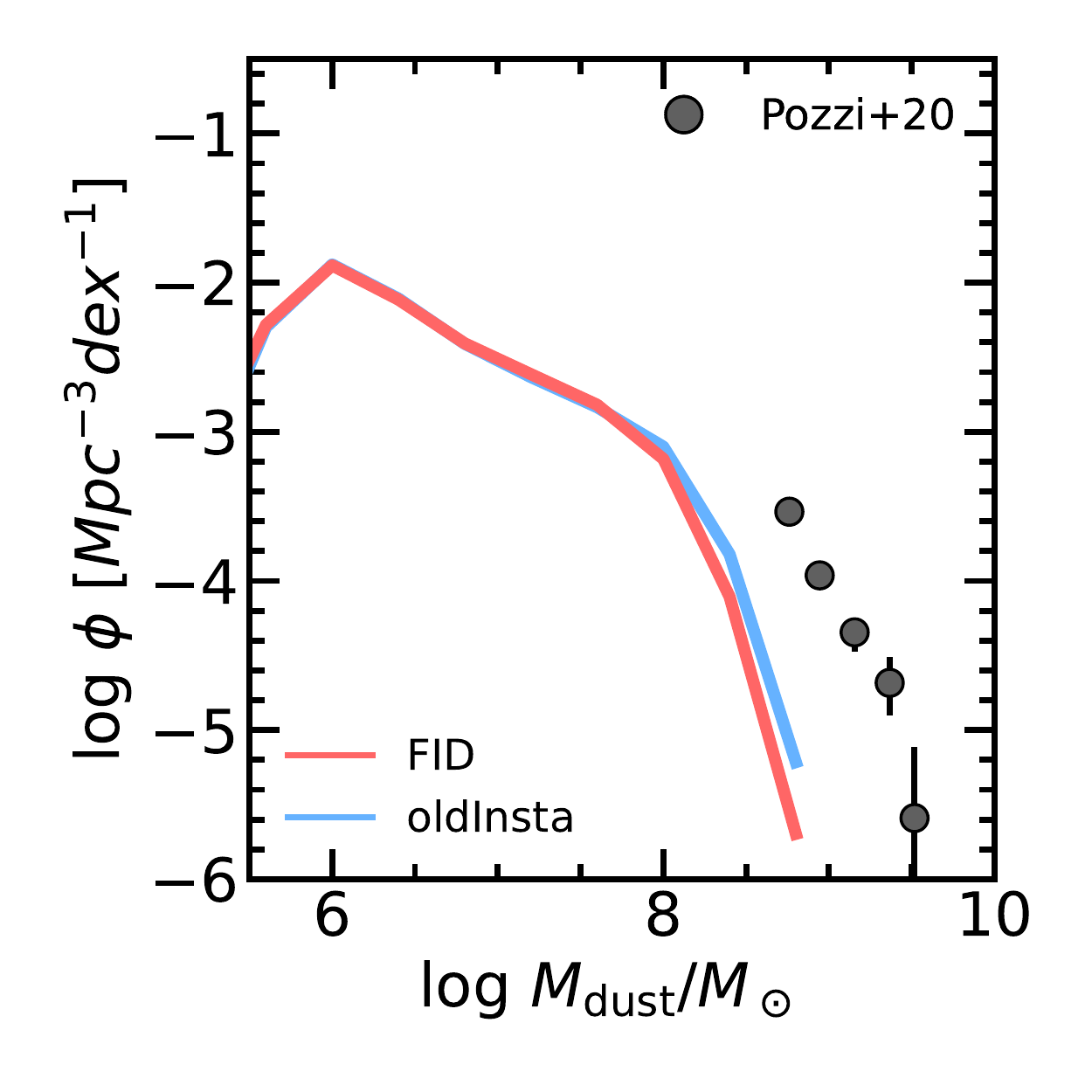}
        \caption{$z=2.25$}
        
    \end{subfigure}
    \caption{Dust Mass Functions of the fiducial model (\textit{FID}; red solid line) and the model without updated disc instability recipe (\textit{oldInsta}; solid line) at $z=0.0, \, 1.0, \, \text{and} \, 2.25$. We also report observations by \citet{Beeston2018}, \citet{Dunne2011} and \citet{Vlahakis2005} at $z=0.0$, and by \citet{Pozzi2020} at higher redshift.}
    \label{fig:DMF}
\end{figure*}

Fig. \ref{fig:DMF} shows the DMF at $z=0.0$, $z=1.0$, and $z=2.25$, together with some observational determinations. The local DMF predicted by our fiducial model is in excellent agreement with the reported observations (\citealt{Vlahakis2005}; \citealt{Dunne2011}; \citealt{Beeston2018}), while the \textit{oldInsta} model overpredicts the high mass end. This failure is strictly related to the HI and H$_2$ mass abundance of the most massive galaxies discussed in Sec. \ref{sec:sanity}, as well as to the number of massive star forming objects illustrated in Fig. \ref{fig:histo-sSFR}. The faster (with respect to the \textit{oldInsta} model) quenching powered by disc instabilities in the \textit{FID} model avoids the presence of very dust rich galaxies, which are not observed today (but instead predicted by other models, see Sec. \ref{sec:dust_cosmic}). In $z>0$ bins, we compare with results by \cite{Pozzi2020}; these authors derived the DMF up to $z \simeq 2.5$ and found a characteristic dust mass increasing with redshift. Qualitatively, also our model predicts more dust rich galaxies in these $z>0$ bins with respect to $z=0$. However, we slightly underestimate the abundance of ${\rm log \,} M_{\rm dust}/M_\odot \gtrsim 9$ high$-z$ galaxies that is observed.

\begin{figure}
    \centering

    \includegraphics[width=0.8\linewidth]{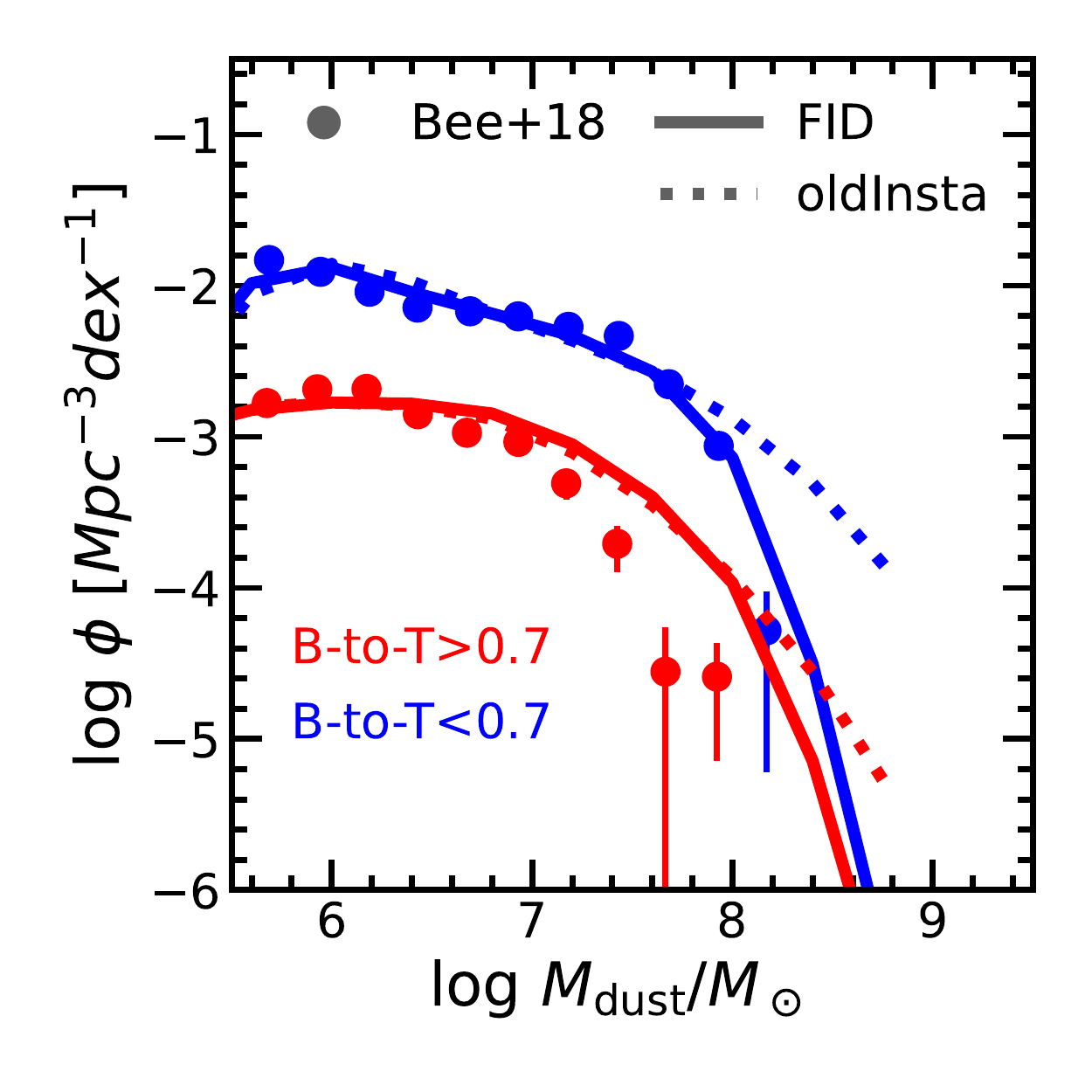} 

    \caption{Dust Mass Function at $z=0$ for elliptical (B-to-T $>0.7$; red) and non-elliptical (B-to-T $<0.7$; blue) galaxies. Results from both the \textit{FID} and \textit{oldInsta} model are shown (solid and dotted lines). We compare with the observations of \citet{Beeston2018}.}
    
    \label{fig:DMFmorpho}
    
\end{figure}

Finally, since our new model of disc instabilities improved the predictions on the morphological type of simulated galaxies (Sec. \ref{sec:morphoSF}), it may be interesting to inspect the relation between the dust content and morphology. \cite{Beeston2018} coupled their dust mass determination with the morphological classification of GAMA galaxies (\citealt{Driver2012}; \citealt{moffett2016}) and derived the local DMF of elliptical and non-elliptical galaxies. We show their results in Fig. \ref{fig:DMFmorpho}, compared with our model predictions. In keeping with data, our elliptical galaxies are more dust poor than non-elliptical ones in both the \textit{FID} and \textit{oldInsta} model. In particular, our inclusion of disc instabilities clearly improves the match for non-elliptical galaxies at the high mass end. The elliptical galaxies DMF is instead less affected by the disc instability prescription.  Both models slightly overestimate the high mass end of their observed DMF.


\subsubsection{Dust, stars and metallicity}

\begin{figure*}
    \centering
    \begin{subfigure}[b]{0.3\textwidth}
        \includegraphics[width=\textwidth]{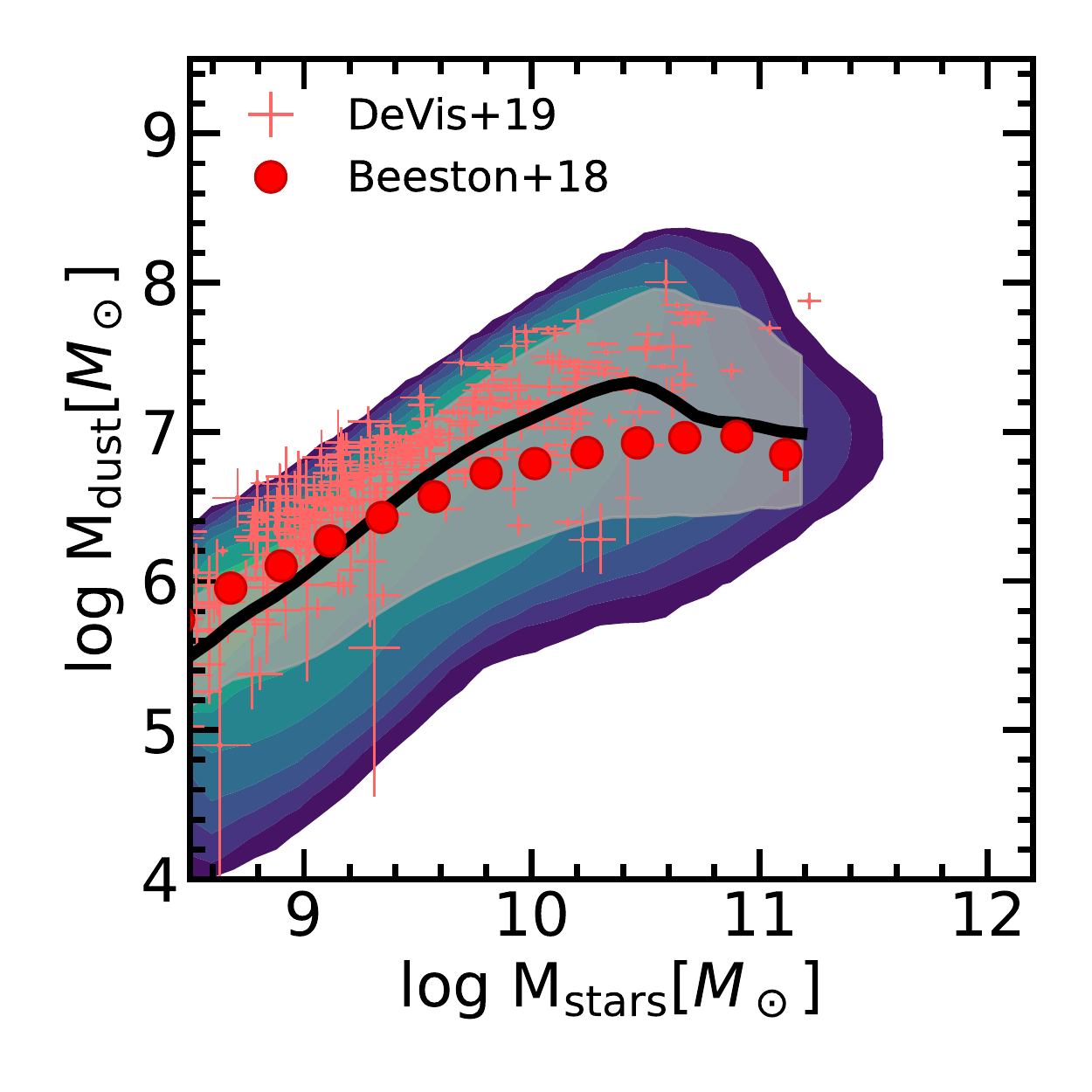}
        \caption{$z=0.0$}
        
    \end{subfigure}
    \begin{subfigure}[b]{0.3\textwidth}
        \includegraphics[width=\textwidth]{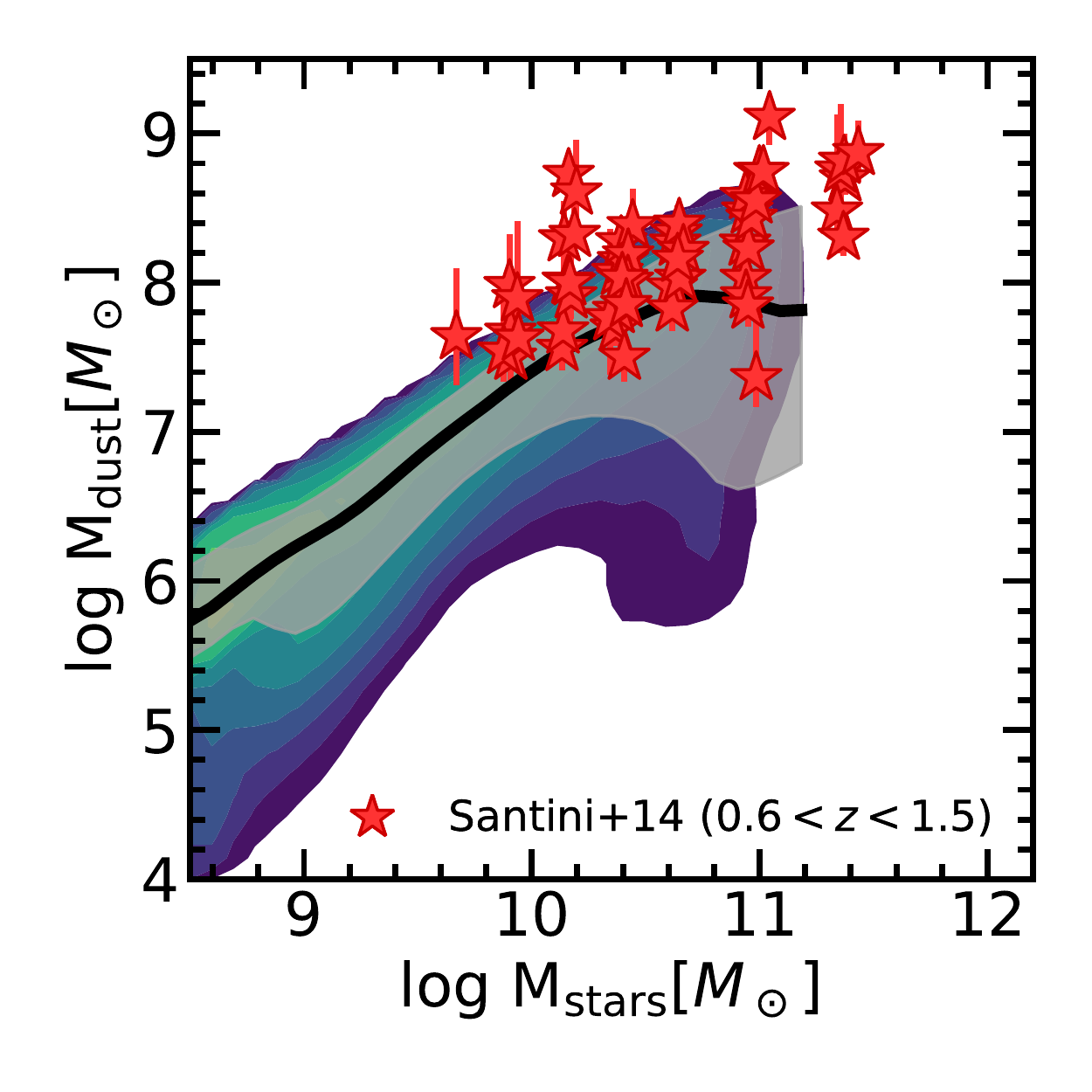}
        \caption{$z=1.0$}
       
    \end{subfigure}
    \begin{subfigure}[b]{0.3\textwidth}
        \includegraphics[width=\textwidth]{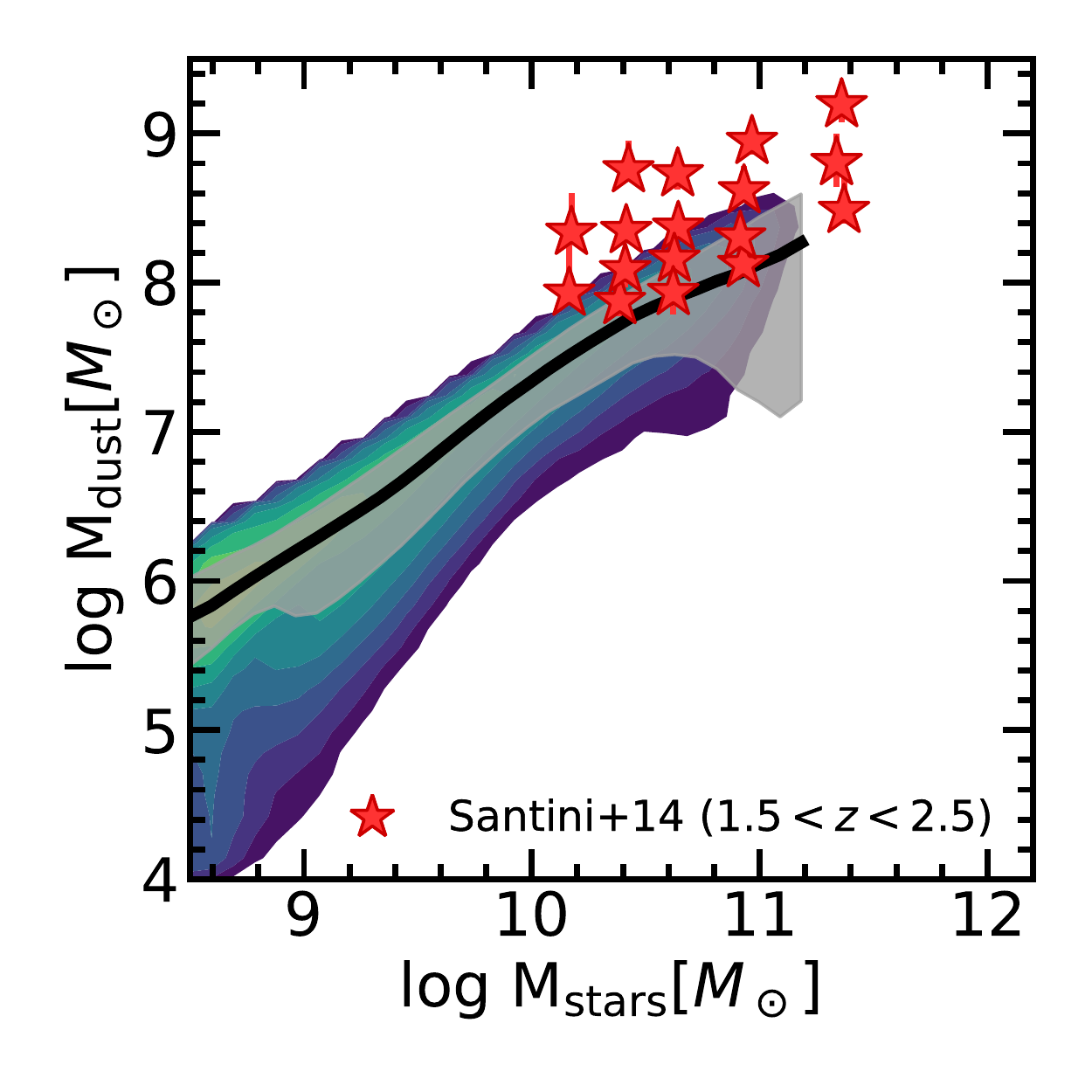}
        \caption{$z=2.25$}
        
    \end{subfigure}
    \caption{$M_{\rm dust}-M_{\rm stars}$ relation for our fiducial model at $z=0.0$, $z=1.0$, and $z=2.25$. The median relation is shown as black line, while the gray shaded region refers to the $16-84$th percentile dispersion. We show in background log-spaced density contours of our galaxies in this plane. Data from \citet{Beeston2018}, \citet{Vis2019}, and \citet{Santini2014} are shown for comparison.}
    \label{fig:Ms-Md}
\end{figure*}

Fig. \ref{fig:Ms-Md} shows the relation between dust and stellar masses in our fiducial model at $z=0.0$, $z=1.0$, and $z=2.25$. Observations from \cite{Beeston2018}, \cite{Vis2019} (DustPedia) and \cite{Santini2014} are used for comparison. Generally, our predictions fit well all data points. In particular, the linear relation observed at $z>0$ redshift bins is reproduced up to $M_{\rm dust} \sim 10^8-10^{8.5} \, M_\odot$. As highlighted in the previous section, galaxies featuring so high dust masses are not observed in the local Universe. At $z=0$ instead, the observed $M_{\rm stars}-M_{\rm dust}$ exhibits a flattening towards large stellar masses (\citealt{Beeston2018}, binned data), as well as our model successfully does. 
                                
\begin{figure*}
    \centering
    \begin{subfigure}[b]{0.3\textwidth}
        \includegraphics[width=\textwidth]{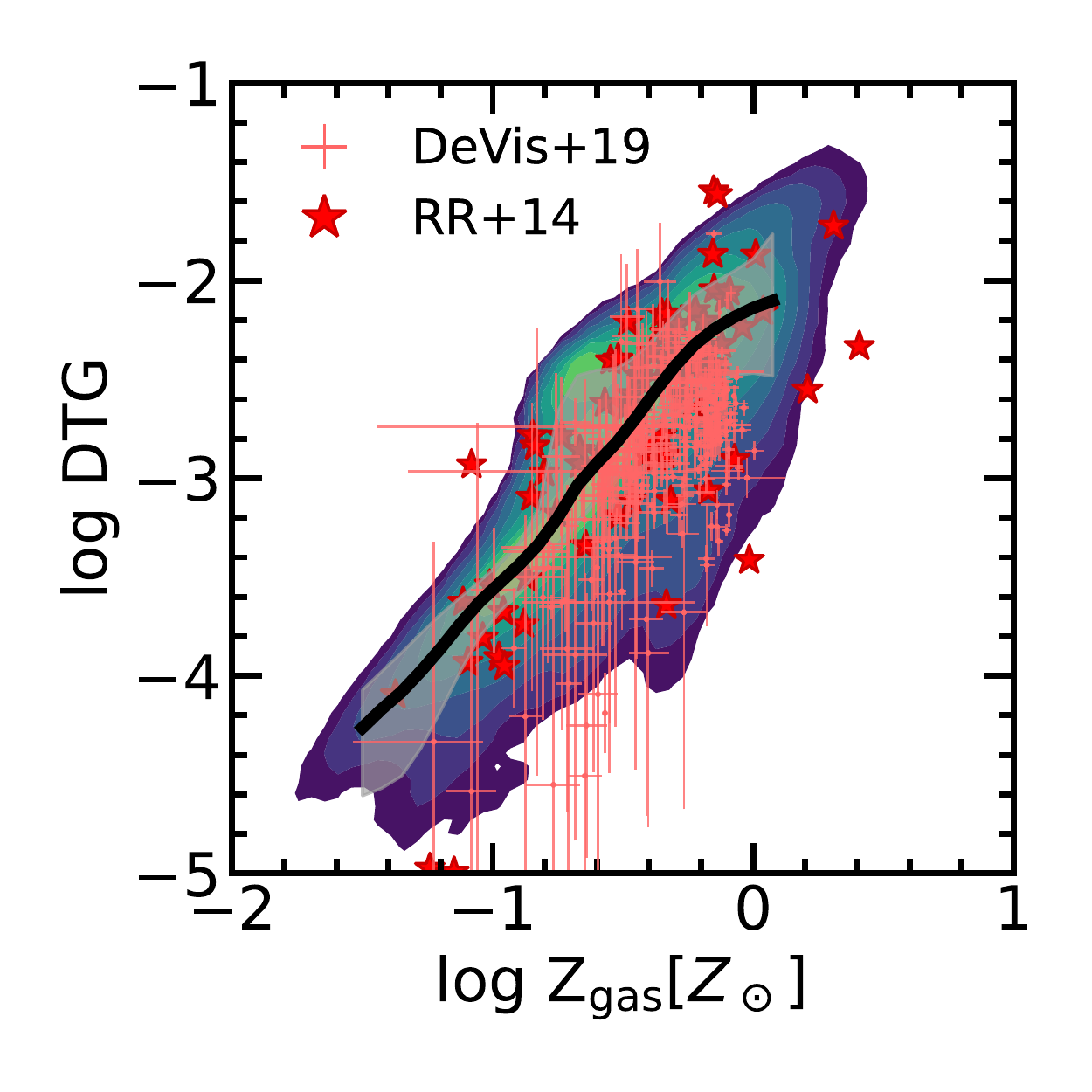}
        \caption{$z=0.0$}
        
    \end{subfigure}
    \begin{subfigure}[b]{0.3\textwidth}
        \includegraphics[width=\textwidth]{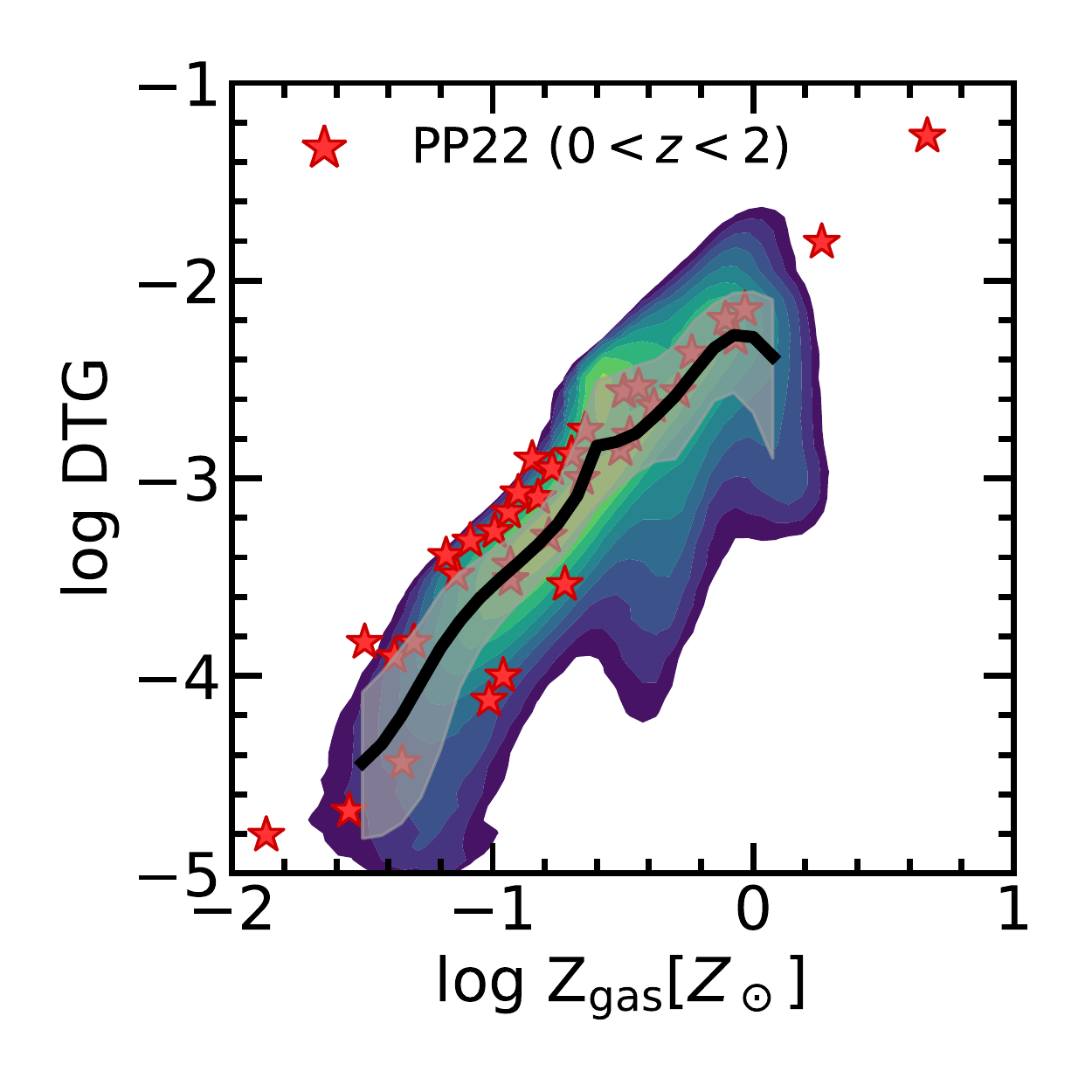}
        \caption{$z=1.0$}
        
    \end{subfigure}
    \begin{subfigure}[b]{0.3\textwidth}
        \includegraphics[width=\textwidth]{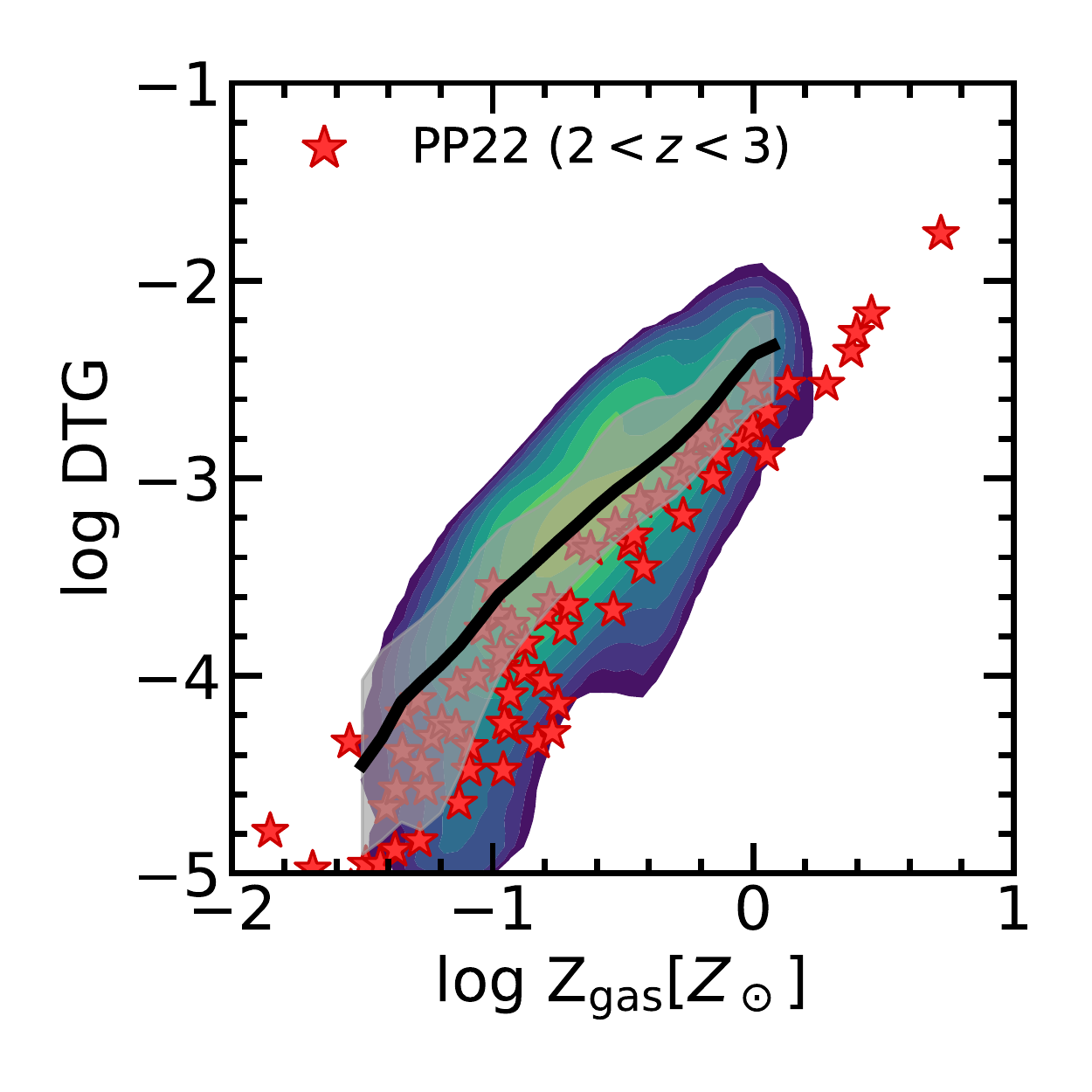}
        \caption{$z=2.25$}
        
    \end{subfigure}
    \caption{DTG$-Z_{\rm gas}$ relation for our fiducial model at $z=0.0$, $z=1.0$, and $z=2.25$. The median relation is shown as black line, while the gray shaded region refers to the $16-84$th percentile dispersion. We show in background log-spaced density contours of our galaxies in this plane. We compare with observations by \citet{Remy-Ruyer2014}, \citet{Vis2019}, and \citet{Popping2022}.}
    \label{fig:DustZ}
\end{figure*}

The relation between ISM Dust-to-Gas ratio (DTG) and metallicity is also crucial since the accretion process relates the two quantities (e.g. \citealt{Hirashita2013}). Thus, we report it in Fig. \ref{fig:DustZ} at various redshifts. Observations (\citealt{Remy-Ruyer2014}; \citealt{Vis2019}; \citealt{Popping2022}) suggest a positive growth of DTG with gas metallicity and a nearly non-evolving relation with cosmic time (see discussion in \citealt{Popping2022}). Our fiducial model reproduces this scenario, especially at $z=0.0$, where different observations well constrain this relation. A nice agreement with data is also obtained at $z>0$, although we slightly overpredict DTG in the $2<z<3$ redshift bin. However, we also point out that no cosmological model currently reproduces this relation over a wide range of redshift; we refer interested readers to \cite{Popping2022} for a discussion of this issue.

\rev{Finally, we note that the lowest metallicity objects in our sample (log $Z_{\rm gas}/Z_\odot \sim -1.5$; $M_{\rm stars}\sim 10^{8.5}\, M_\odot$) feature a very small, but \textit{non-zero}, amount of dust, i.e. $M_{\rm dust}\sim 10^4 \, M_\odot$. These objects, often referred to as extremely metal-poor galaxies, are also observed to be dust deficient (e.g. \citealt{Almeida2016}) and are thought to be analogous of high-$z$ galaxies (e.g. \citealt{Fisher2013}; but see also \citealt{Isobe2021}).}

\subsubsection{Grain sizes}

\begin{figure*}
    \centering
    \begin{subfigure}[t]{0.35\textwidth}
        \centering
        \includegraphics[width=\linewidth]{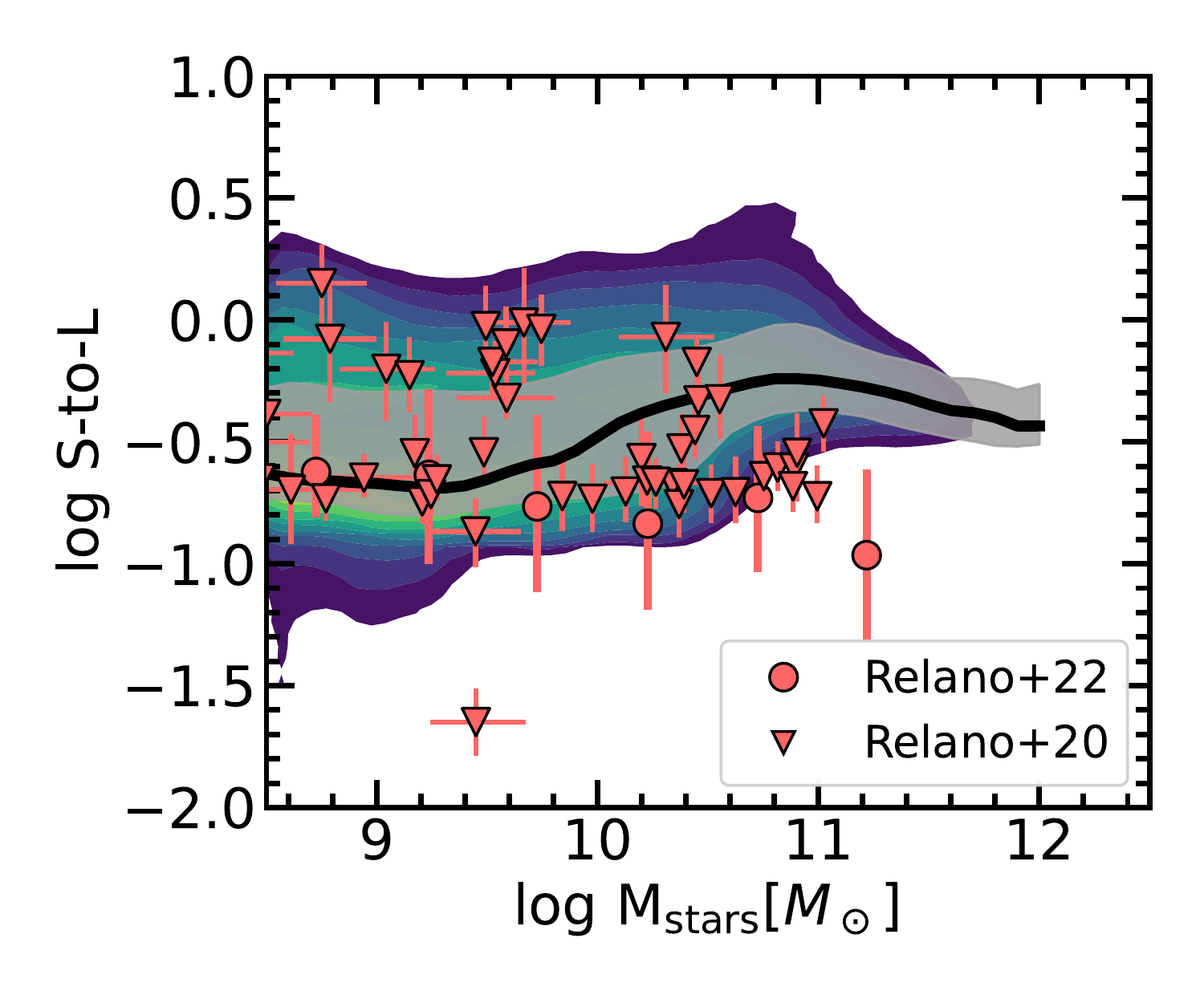}
        \caption{}
        
    \end{subfigure}
    \quad
    \begin{subfigure}[t]{0.35\textwidth}
        \centering
        \includegraphics[width=\linewidth]{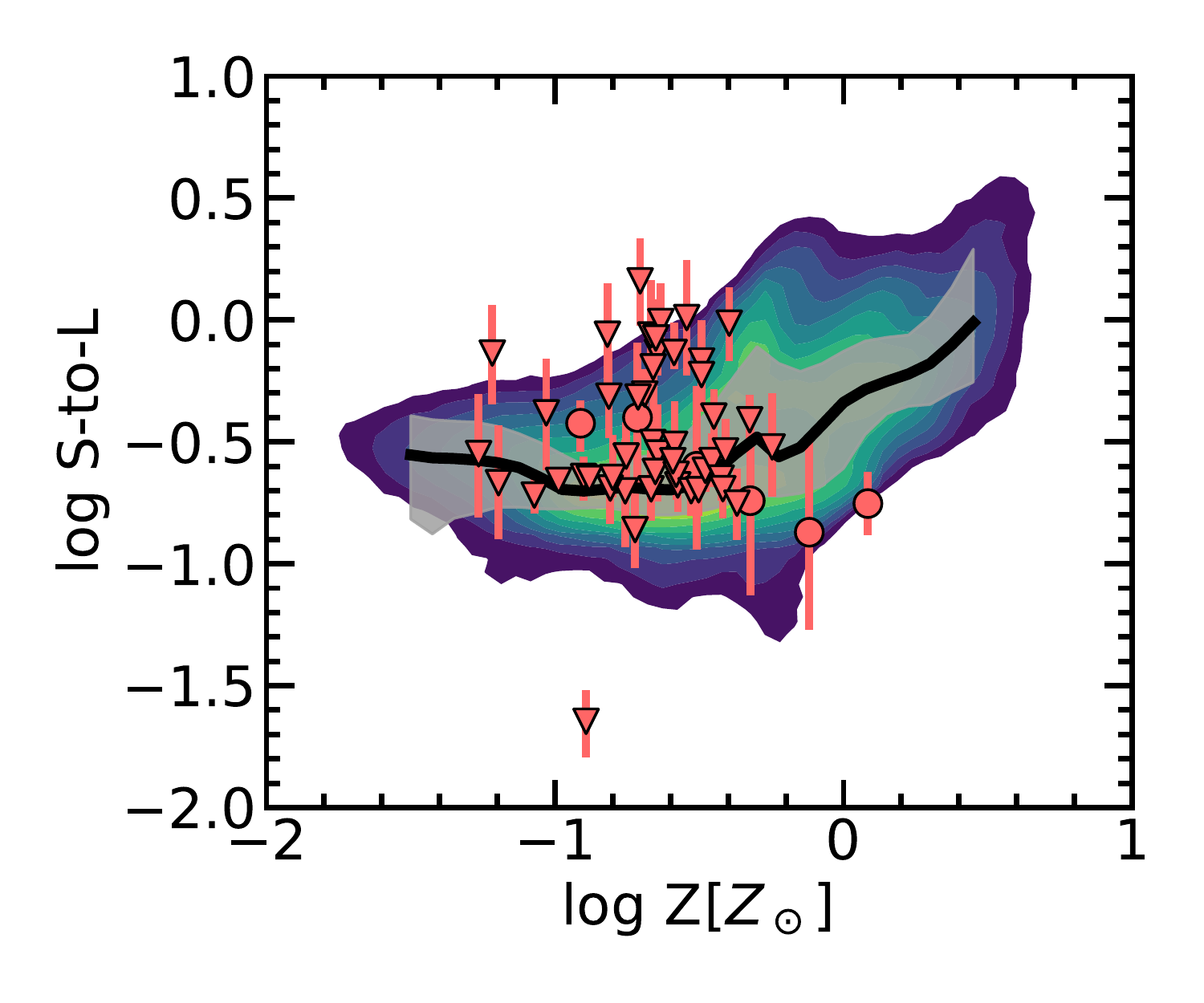} 
        \caption{}
        
    \end{subfigure}

    \caption{Small-to-Large grains mass ratio as a function of stellar mass (\textit{left panel}) and ISM total metallicity (\textit{right panel}) for our fiducial model (density contours) at $z=0.0$. The median relation is shown as black line, while the gray shaded region refers to the $16-84$th percentile dispersion. We show in background log-spaced density contours of our galaxies in this plane. The SAM results are compared with observations by \citet{Relano2020} (red triangles; both DSG and KINGFISH data), and \citet{Relano2022} (red circles; binned data).}
    \label{fig:SL}
    
\end{figure*}

Our dust model follows two distinct dust grain sizes, i.e. small and large grains. Observationally, the determination of the relative abundance of small and large grains may be estimated with SED fitting. This estimate has been done for a sample of local galaxies in two recent works by \citet{Relano2020} and \citet{Relano2022}, whose results are shown in Fig. \ref{fig:SL} together with our model predictions. We caution the reader about this comparison. Indeed, while in our two size approximation the transition from small to large grains occurs at $0.03 \, \mu\text{m}$, it occurs at $0.015 \, \mu\text{m}$ in the model\footnote{The adopted cosmic dust model is that by \cite{Desert1990}. It takes into account polycyclic aromatic hydrocarbons (PAHs), very small grains (VSGs), and big grains (BGs). The former two are considered small grains, the latter component is the only constituent of large grains.} adopted by \cite{Relano2020} and \cite{Relano2022}. Moreover, they assume carbonaceous properties for small grains, while we also consider small grains to have a silicate composition.

The relation between the observed Small-to-Large grain mass ratio (S-to-L) and stellar mass (left panel) exhibits e nearly flat trend with a median value of S-to-L$\sim 0.2-0.3$ and $\sim 0.7$ dex dispersion. However, the more recent work by \cite{Relano2022}, who made use of a significantly larger sample of galaxies, suggests a slowly decreasing trend of S-to-L with $M_{\rm stars}$. The relationship between S-to-L and ISM metallicity (right panel) shows qualitatively similar behaviour. Although our model matches the median S-to-L, we fail in reproducing its decrease toward high $M_{\rm stars}$ and $Z$. Other cosmological simulations which adopt the two size approximation (\citealt{Hou2019}; \citealt{Parente2022}) ascribe this behaviour to the dominance of coagulation over accretion and shattering. In our model, high S-to-L galaxies at large $M_{\rm stars} \, \text{or} \, Z$ have typically low molecular gas fractions. Thus coagulation of small grains is not effective enough. We conclude that, although our model can reproduce the S-to-L ratio of the bulk of the observations, some improvements of the admittedly simple treatment of shattering and coagulation are required to avoid over-predicting the abundance of small grains in massive galaxies.
Finally, we note that the adoption of a non-constant $n_{\rm gas}$ in Eq. \ref{eq:shat} (see also Eq. \ref{eq:rhoshat}) has little impact on the scaling relations just discussed. Instead, it is needed to obtain reasonable results when looking at small and large grain profiles, as we discuss in Appendix \ref{app:SLprofile}. \\


\subsection{The cosmic evolution of dust abundance}
\label{sec:dust_cosmic}
In this section we present and discuss what we consider the most interesting result of our work, that is the cosmic evolution of the galactic, extra-galactic and total dust budget across cosmic time.

\subsubsection{Galactic dust budget}
\label{sec:Omega_gal}

\begin{figure*}
    \centering

    \includegraphics[width=0.8\linewidth]{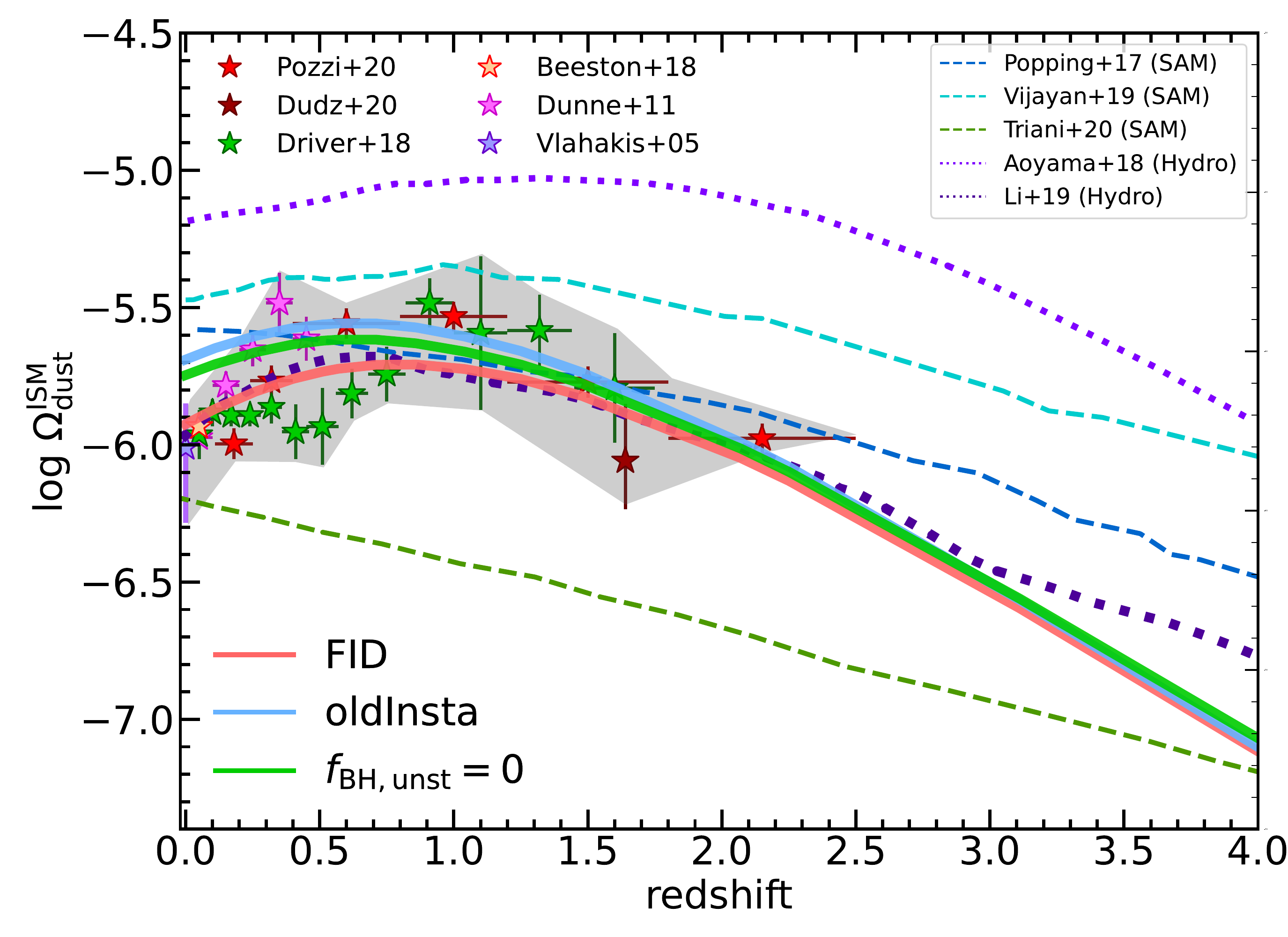} 

    \caption{Cosmic ISM dust abundance across cosmic time. We show the dust abundance in cold gas for: \textit{(i)} our fiducial model (\textit{FID}; red solid line); \textit{(ii)} a model without the new implementation of disc instabilities discussed in this work (\textit{oldInsta}; blue solid line); \textit{(iii)} a model without BH accretion during disc instabilities ($f_{\rm BH, unst}=0$ in Eq. \ref{eq:BHaccinsta}; green solid line). Results derived from the integration of the DMF at various redshifts are shown as stars: \citet{Vlahakis2005} (purple), \citet{Dunne2011} (pink), \citet{Beeston2018} (orange), \citet{Driver2018} (green), \citet{Dudzeviciute2020} (brown), and \citet{Pozzi2020} (red). A shaded grey region shows the global behaviour of these data. Results from other groups are shown as dashed (SAMs; \citealt{Popping2017} in blue, \citealt{Vijayan2019} in light cyan, \citealt{Triani2020} in green) and dotted (hydrodynamic simulations; \citealt{Aoyama2018} in purple, \citealt{Li2019} in dark blue) lines.}
    \label{fig:rhodust}
    
\end{figure*}

The galactic dust mass density $\Omega^{\rm ISM}_{\rm dust}$ as a function of redshift for the two models discussed above are shown in Fig. \ref{fig:rhodust}, together with some observational determinations and predictions from other cosmological models. We also show the result obtained with a model in which $f_{\rm BH, unst}=0$ (Eq. \ref{eq:BHaccinsta}), that is we do not allow for BH growth during disc instabilities. We compare with the observations by \cite{Vlahakis2005}, \cite{Dunne2011}, \cite{Beeston2018},  \cite{Driver2018}, \cite{Dudzeviciute2020}, and \cite{Pozzi2020}, which are obtained by integrating the observed DMF.

Both the \textit{FID} and \textit{oldInsta} models are in agreement with data within a factor $\sim 2$. While the model without the new implementation of disc instabilities reproduces better $z \gtrsim 1$ data, our fiducial model performs better at lower redshift, in particular as far as the observed decrease of $\Omega^{\rm ISM}_{\rm dust}$ is concerned. As already discussed, the different performance of these two models is strictly related to the BH accretion during disc instabilities and its effect on quenching. This is confirmed by the $f_{\rm BH, unst}=0$ model, which performs similarly to the \textit{oldInsta} model and does not reproduce the desired decline of $\Omega^{\rm ISM}_{\rm dust}(z)$ toward $z=0$. 

Fig. \ref{fig:rhodust} demonstrates that the observed $\Omega^{\rm ISM}_{\rm dust}(z)$ is poorly reproduced by almost all other cosmological computations (SAM or hydrodynamical simulations) published so far, which tend to overpredict the data and to produce only a very mild, if any, decrease from $z\sim 1$ to $z=0$. The only exception is the SIMBA simulation (\citealt{Li2019}), which shows an $\Omega^{\rm ISM}_{\rm dust}$ evolution very similar to that of our fiducial model.\\

\begin{figure}
    \centering

    \includegraphics[width=1.\linewidth]{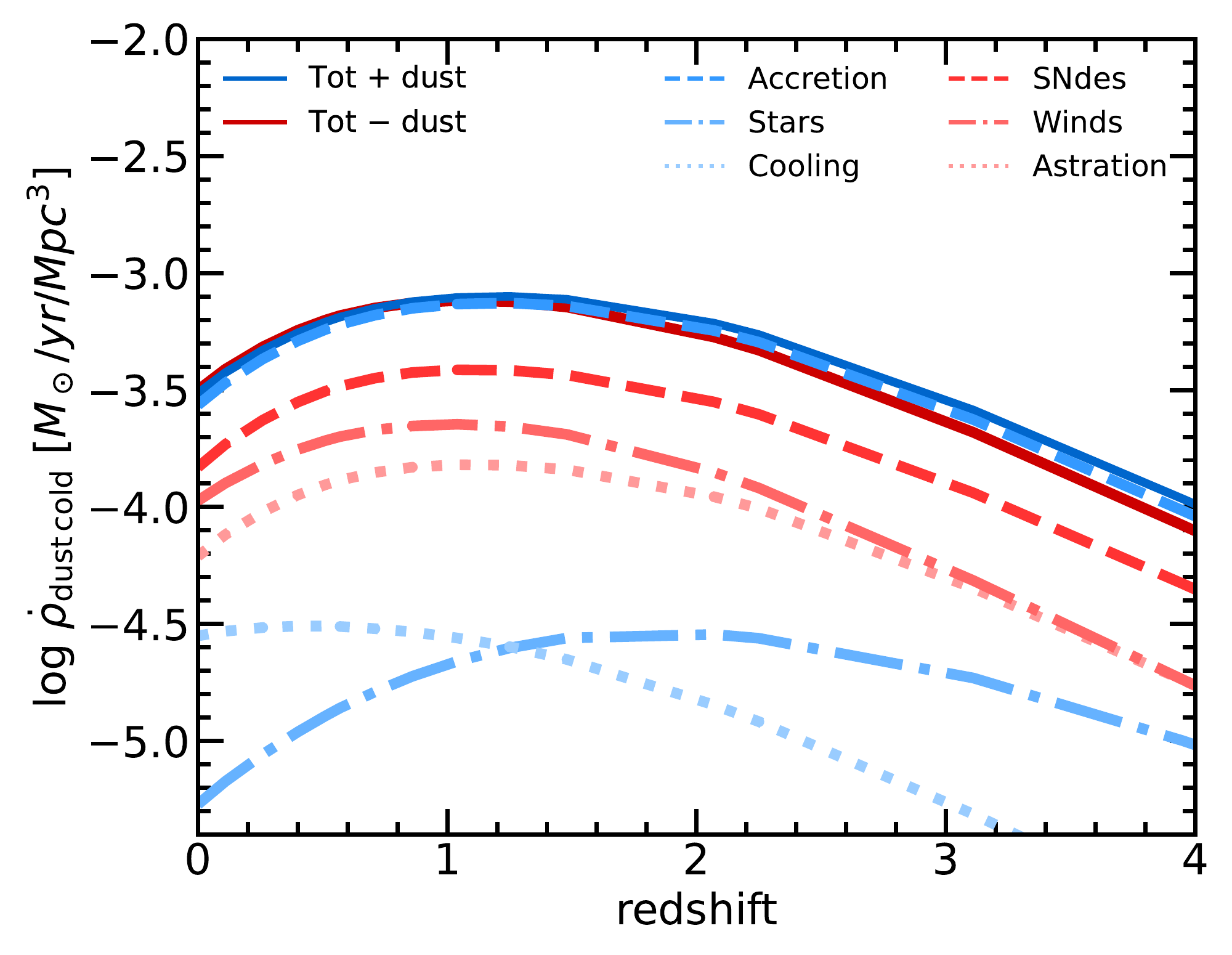} 

    \caption{Cosmic evolution of the rates of the processes contributing to the overall abundance of dust in the cold gas of galaxies in our fiducial model: stellar production (blue dot-dashed), accretion (blue dashed), dust transferred from hot to cold gas by cooling (blue dotted), destruction by SN (red dashed), dust ejected from cold by stellar driven winds (red dot-dashed), and astration (red dotted). The total rate of the processes leading to an increase(decrease) of dust in the cold gas is shown as a solid blue(red) line.}
    \label{fig:dustrates}
    
\end{figure}

We now investigate \textit{why} and \textit{how} the galactic dust decreases from $z \sim 1$ to $z=0$ in our fiducial model. To do this, we consider separately the rates of all the processes capable of modifying the abundance of the dust in the cold gas of our model galaxies: production by stars, accretion, dust transferred from the hot to the cold gas by cooling, destruction by SNe, ejection from cold gas, and astration. These are shown in Fig. \ref{fig:dustrates} as a function of redshift, together with the sum of those contributing to a decrease(increase) of $\rho_{\rm dust\, cold}$. Except for dust \textit{cooled} (i.e. dust transferred from the hot to cold gas in the cooling process), all other rates roughly follow the shape of the SFRD with a certain delay. This result is expected since all of them depend directly (dust produced by stars, astration) or indirectly (destruction by SN, ejection from cold gas due to stellar feedback, accretion) on star formation. Among the processes contributing to the increase of dust, accretion is by far the dominant one ($\gtrsim 80-90\%$ at least for $z \lesssim 4$, where we focus our attention). On the contrary, all  destruction/ejection processes have a significant contribution to the decrease of ISM dust (destruction by SN $\simeq 45 \%$, ejection $\simeq 35 \%$ and astration $\simeq 20 \%$ at $z=0$).

\begin{figure}
    \centering
    \includegraphics[width=1.\linewidth]{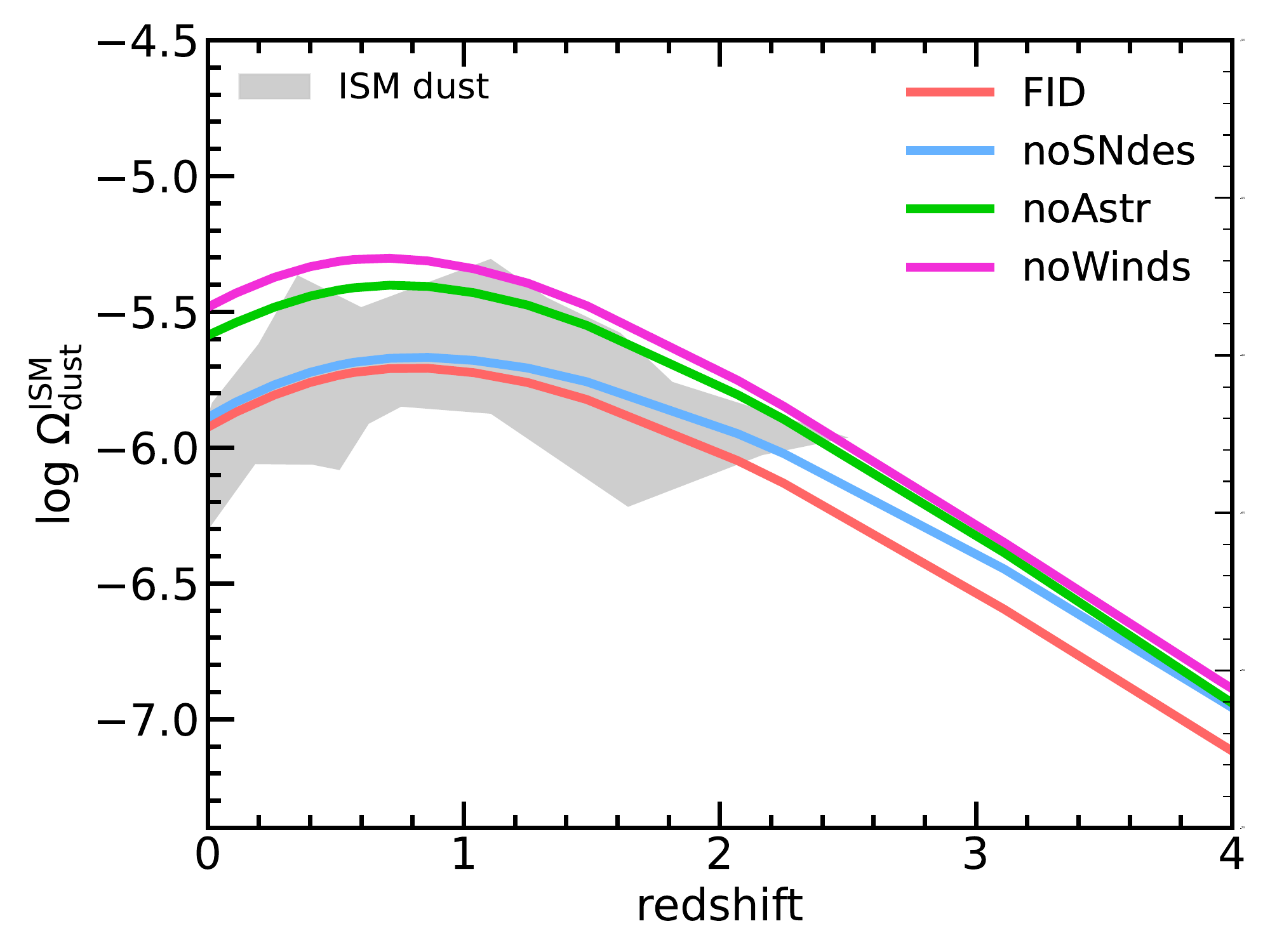} 

    \caption{Cosmic evolution of galactic dust obtained by switching off one at a time the processes responsible for the dust destruction/removal in the cold gas: destruction by SNae (\textit{noSNdes}; blue), astration (\textit{noAstr}; green), ejection from cold gas (\textit{noWinds}; pink). We also report for comparison our fiducial (\textit{FID}; red) model and the compilation of observations introduced in Sec. \ref{sec:Omega_gal} (gray shaded region).}
    \label{fig:dustproc}
    
\end{figure}

An interesting exercise is to study what happens to $\Omega^{\rm ISM}_{\rm dust}$ when the destruction/ejection processes (astration, ejection from cold gas, and SN destruction) are alternatively switched off.
Fig. \ref{fig:dustproc} shows the result. When neglecting dust destruction by SNe (\textit{noSNdes}), the predicted $\Omega^{\rm ISM}_{\rm dust}$ is very similar to our fiducial model, but for a slight increase of it, especially at high $z$. This behaviour is because most of the gas phase metals produced by the SNe destruction process in the \textit{FID} model are quickly re-locked into dust grains by the accretion process. Therefore, in the \textit{noSNdes} model, this does not happen, also accretion is reduced with respect to the \textit{FID} model, and the final $\Omega^{\rm ISM}_{\rm dust}$ does not differ much. Thus, the decline of $\Omega^{\rm ISM}_{\rm dust}$ is present also in this model without SN destruction. 

As for the importance of SN destruction, \cite{Ferrara2021} recently argued that the observed $\Omega_{\rm dust}$ $z 
\sim 1 \rightarrow 0$ drop may suggest that each SN destroys $M_{\rm des,SN}\simeq 0.45 \, M_\odot$ of dust, much lower than $M_{\rm des,SN}\simeq 4.3 \, M_\odot$ that, according to them, is commonly adopted. \cite{Ferrara2021} obtain it assuming a solar DTG ($\sim 0.006$). Actually, in our computations we have $M_{\rm des,SN}\simeq 2.5 \, M_\odot$ for galaxies with DTG$\simeq$DTG$_\odot$, but when considering the whole galaxies population we obtain a median $M_{\rm des,SN}\simeq 0.71(0.47) \, M_\odot$ at $z=0(1)$, similar to the value proposed by \cite{Ferrara2021}. 
We also confirm that $M_{\rm des,SN}\simeq 4.3 \, M_\odot$ would be not acceptable. Indeed, we checked that this median value is obtained by increasing the SN destruction efficiency by a factor $\sim 6-10$, and it under-predicts (by a factor $\sim 1.5-2.5$) $\Omega^{\rm ISM}_{\rm dust}$ at all redshift. The accretion process is not efficient enough to re-lock all the metals destroyed by SNe into dust.

As for the other destruction/ejection processes, a similar decline of $\Omega^{\rm ISM}_{\rm dust}$ occurs when astration and ejection from cold gas are switched off, but for the normalization, which is higher than in the fiducial model, since we are shutting down channels able to remove dust grains and the metals locked in them from the cold gas. Note that this is intrinsically different from what SNe destruction does. Indeed, the latter process transfers dust metals to gas metals, still leaving them available for new accretion processes in the cold phase.\\

We conclude that the dust processes in our model provide a good normalization of $\Omega^{\rm ISM}_{\rm dust}$, and we have shown that none is the only one responsible for its decline. The cosmic evolution of galactic $\Omega^{\rm ISM}_{\rm dust}$ is intrinsically linked to the star formation history (and thus to the neutral gas and molecular abundance evolution in galaxies), since all processes depend on it to some degree. Our new treatment of disc instabilities and the associated SMBH growth reduces the star formation, gas, and dust content of massive local galaxies, sharpening the $\Omega^{\rm ISM}_{\rm dust}$ drop towards $z=0$, which improves the match with low redshift data.


\subsubsection{Extra-galactic dust budget}
\label{sec:Omega_hot}
\begin{figure}
    \centering

    \includegraphics[width=1.\linewidth]{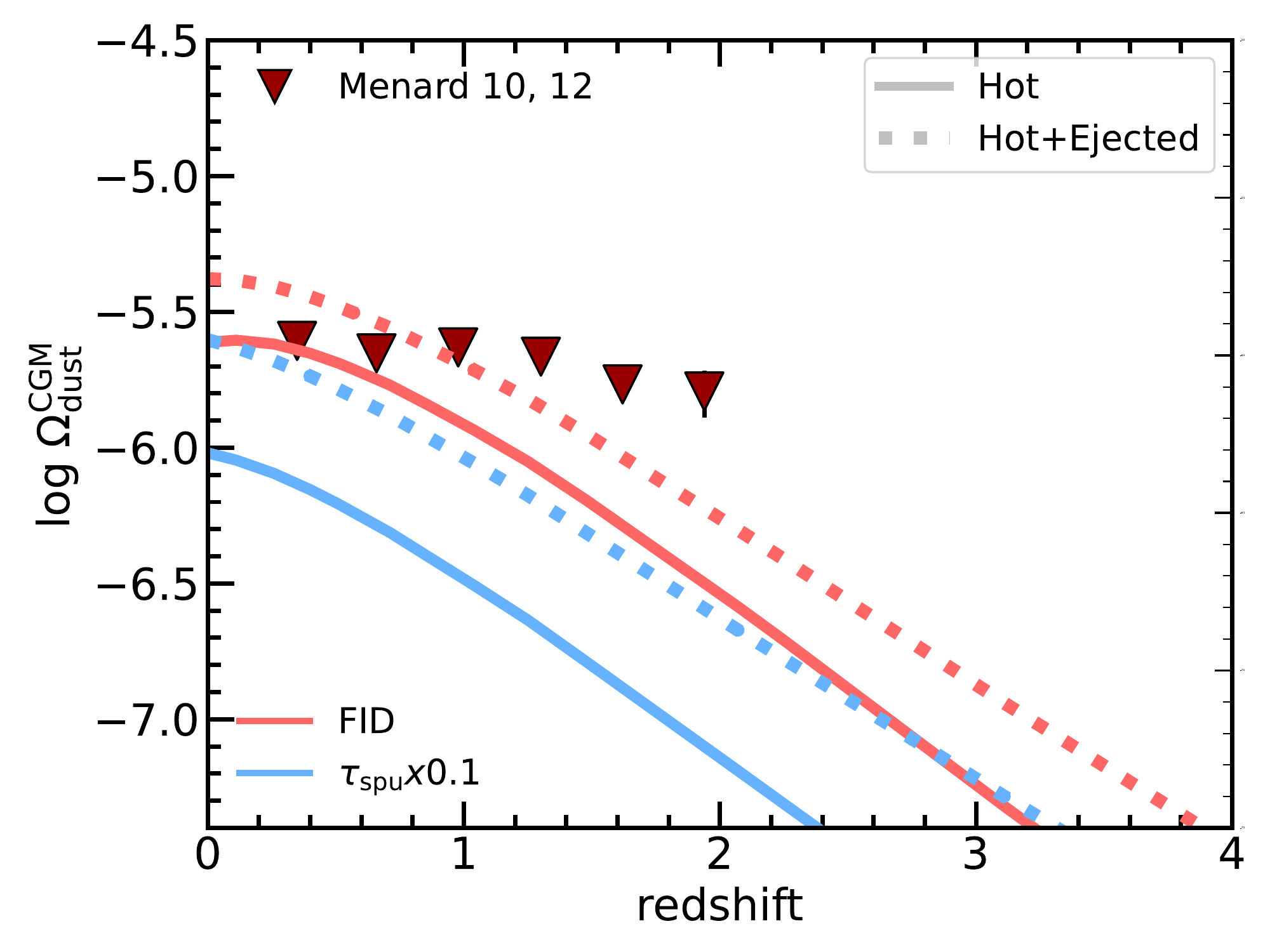} 

    \caption{Cosmic evolution of dust in hot gas (solid line) and both hot gas and ejected reservoir (dotted line). The fiducial model is shown as red lines, while a model with a sputtering efficiency enhanced by a factor $10$ (i.e. adopting the timescale originally introduced by \citealt{Tsai1995}) is shown as blue lines. We report observations by \citet{Menard2010} (the lowest redshift point) and \citet{Menard2012}, corresponding to CGM dust.}
    
    \label{fig:OmegaDustCGM}
    
\end{figure}

We now discuss our model predictions concerning the amount of extra-galactic dust. As anticipated in the Introduction, we have just a couple of observational determinations of the dust residing beyond the ISM of galaxies, namely the ones derived by \cite{Menard2010} and \cite{Menard2012}. In Fig. \ref{fig:OmegaDustCGM} we compare their determinations of $\Omega^{\rm CGM}_{\rm dust}$ with the dust residing in the hot halo of our model galaxies, which may be broadly identified with the CGM of galaxies. However, in the SAM framework, there exists also another extra-galactic field, namely the ejected reservoir. This component represents a reservoir of hot gas that can not cool (differently from \textit{hot gas}), but its physical interpretation is ambiguous. For this reason, we also report the extreme case of considering CGM dust that resides in both the hot gas and ejected reservoir. We recall that in our fiducial model, we adopt a sputtering efficiency lower by a factor of $10$ than the one originally introduced by \cite{Tsai1995}, as we anticipated in Sec. \ref{sec:dustspu}. This is in keeping with results by  \cite{Gjergo2018} and \cite{Vogelsberger2019}, from computations of dust evolution in zoom-in simulation of galaxy clusters. They found that the sputtering efficiency by \cite{Tsai1995} leads to a too low dust content in the ICM, with respect to recent observations based on Planck data \citep{PlanckCollaborationXLIII2016}.
In this section we provide additional evidence in favor of a lower efficiency by showing also results obtained by a model dubbed $\tau_{\rm spu}{\rm x}0.1$ which adopts the \cite{Tsai1995} sputtering efficiency.

When considering hot dust alone, the latter model underestimates the observations at all cosmic times by a factor $3-10$, with a tension decreasing towards low $z$. The hot+ejected dust is instead in good agreement with observations at the lowest redshifts, but a tension of a factor $\sim 5$ still exists at $z\sim 2$. Contrarily, adopting a longer sputtering timescale as in the fiducial model helps to reproduce a CGM dust abundance more in line with observations, especially when considering both hot and ejected dust. We also note that the available observations do not allow to conclude that extra-galactic dust presents an abundance drop at $z \lesssim 1$, as observed for ISM dust. Instead, observations suggest a nearly flat trend with redshift, in tension with our steeper results; this may indicate that we are missing some physics. Indeed, although sputtering is the only dust process directly affecting the amount of dust in the hot gas (and in the ejected reservoir), this quantity also depends on several physical processes implemented in the SAM. For instance, SN driven winds are the main dust enrichment channel of the hot gas. Assuming an ejected material with a different cold gas DTG may affect our results, as well as the implementation of AGN-driven outflows, currently lacking in this model. \\

\begin{figure}
    \centering    
    \includegraphics[width=0.9\linewidth]{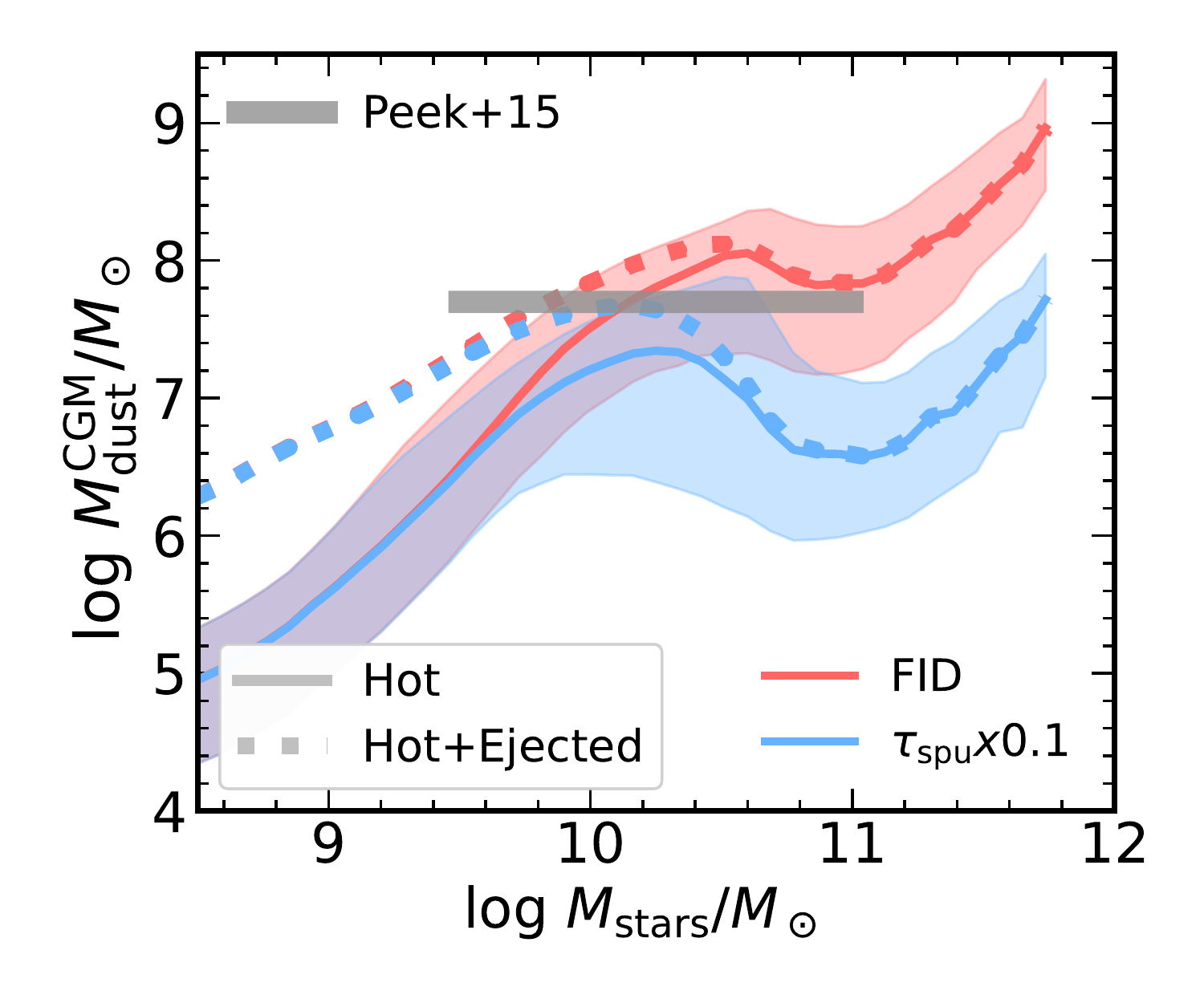} 

    \caption{CGM dust as a function of stellar mass when considering dust in the hot gas (solid, with $16-84$th percentiles dispersion) and both hot and ejected gas (dotted). The fiducial model is shown as red lines, while a model with a sputtering efficiency reduced by a factor of $10$ is shown as blue lines. We report results by \citet{Peek2015} as a gray line.}
    
    \label{fig:MdustCGM}
    
\end{figure}

The good agreement obtained at low$-z$ with a reduced sputtering efficiency is confirmed by the comparison with the work by \cite{Peek2015}. These authors studied the CGM reddening of low$-z$ galaxies, deriving typical CGM dust masses of $\simeq 6 \cdot 10^7 \, M_{\odot}$ and a weak dependence on stellar mass for $0.1L_*-L_*$ galaxies. We thus report their results in Fig. \ref{fig:MdustCGM} together with our predictions for both the fiducial and the $\tau_{\rm spu}\rm{x}0.1$ model. The \textit{FID} model allows us to capture CGM dust values found by \cite{Peek2015}, and we note that considering both the hot gas and ejected reservoir as CGM dust improves the comparison with the observed, nearly flat slope of this relation.

From the comparison of the \textit{FID} and $\tau_{\rm spu}\rm{x}0.1$ models, it is clear that sputtering has a relevant role only above a certain stellar mass $M_{\rm stars}\gtrsim 10^{10}\, M_\odot$, where virial temperatures are high enough to make this process efficient. Moreover, in both models, ejected dust has a relevant contribution only below the aforementioned critical mass, above which SN winds are no longer efficient in making gas (and thus dust) able to escape from the hot halo. These qualitative findings are in keeping with what has been shown and discussed by \cite{Popping2017} and \cite{Triani2020}, although within different SAM frameworks. Finally, we note that above $M_{\rm stars} \gtrsim 10^{11} \, M_\odot$ the hot phase dust increases, since $T_{\rm vir}\gtrsim T_{\rm spu,0}$, that is the temperature above which the sputtering efficiency saturates (see Eq. \ref{eq:tauspu}). \\

To conclude this section, we remark that the admittedly scanty information on the extra-galactic dust budget requires a lowering of the thermal sputtering efficiency, as already noted by \cite{Gjergo2018} and \cite{Vogelsberger2019} for hydro-dynamic simulations of galaxy clusters. However, we caution about the comparisons performed here since CGM dust in a SAM framework is somewhat not well defined. 


\subsubsection{Total dust budget}

\begin{figure}
    \centering

    \includegraphics[width=1.\linewidth]{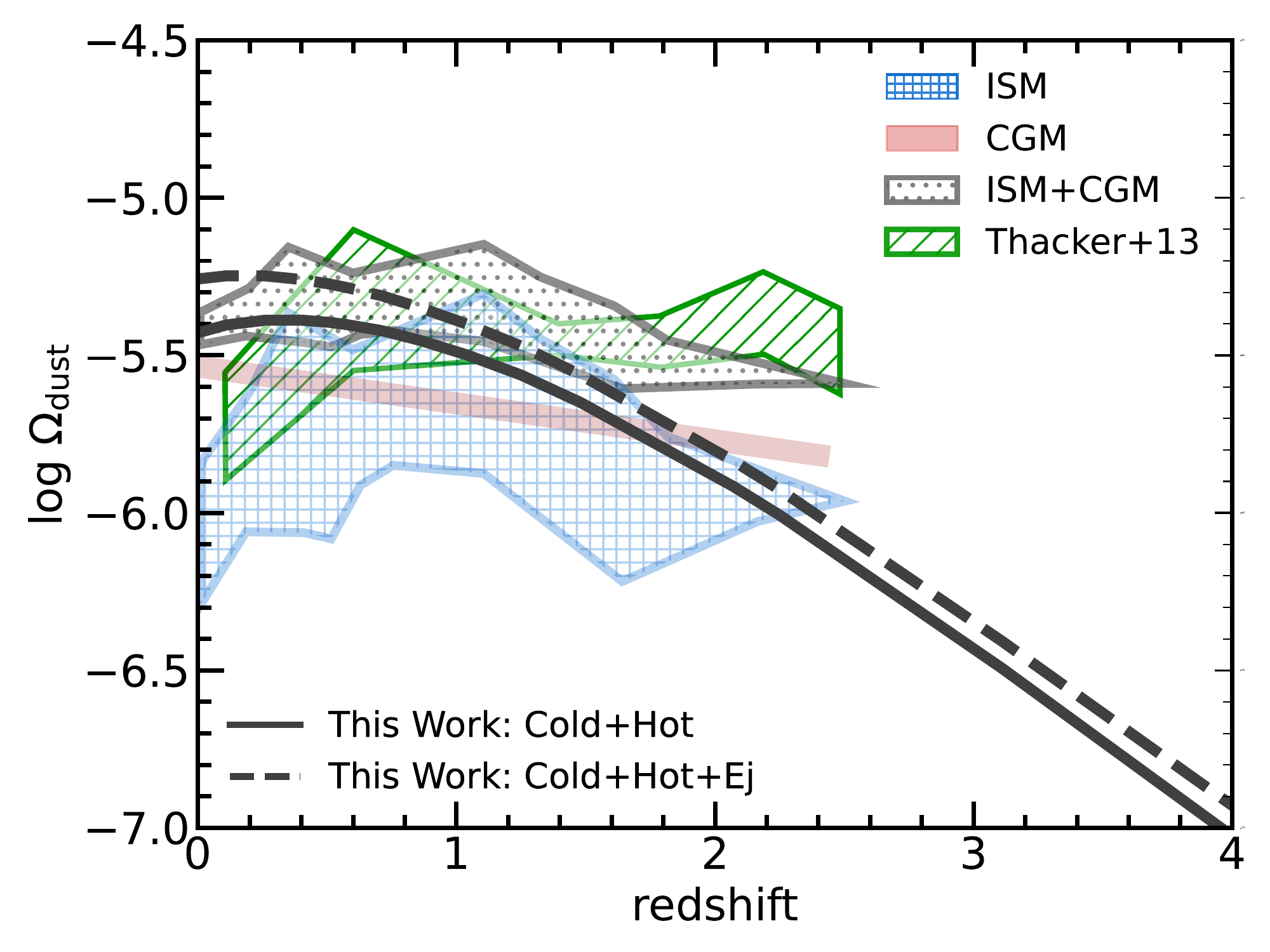} 
    \caption{
    \rev{Cosmic evolution of the total dust budget (Cold+Hot, black solid line, and Cold+Hot+Ej, black dashed line) as predicted by our fiducial model. Shaded and hatched regions correspond to different sets of observations. ISM dust (blue hatched region) refers to the compilation of observations shown in Section \ref{sec:Omega_gal}. CGM dust (red shaded region) is obtained by performing a linear fit, extrapolated down to $z=0$, of \citet{Menard2010} and \citet{Menard2012} data (Sec. \ref{sec:Omega_hot}). Summing the above components (ISM+CGM) we obtain an estimate of the total dust content (gray hatched region). Results obtained by \citet{Thacker2013} are shown as a green hatched region. }
    }
    
    \label{fig:Omegatot}
    
\end{figure}

In this section, we put together the results concerning the amount of galactic and extra-galactic dust and thus discuss the cosmic evolution of the total amount of dust in our simulation. Deriving an accurate observational estimate of this quantity is a challenging task. Thus, we consider two different estimates. The first one is simply the sum of the galactic ISM ($\Omega^{\rm ISM}_{\rm dust}$, that is the compilation of observations introduced in Sec. \ref{sec:Omega_gal}) and CGM contribution\footnote{To take into account the contribution of CGM observations, we perform a linear fit extrapolated down to $z=0$. We have no reasons to assume something more complex than this.} ($\Omega^{\rm CGM}_{\rm dust}$, \citealt{Menard2010} and \citealt{Menard2012} data, Sec. \ref{sec:Omega_hot}). The second one is the result of \cite{Thacker2013}, obtained with the observation of FIR background. This technique should capture the thermal emission of the whole population of dust grains in our Universe. 
Both these determinations are shown in Fig. \ref{fig:Omegatot}, together with our model predictions. Moreover, we show results considering the CGM made either by hot gas only or hot plus ejected gas. 

First, we note that our naif derivation of the total amount of dust and \cite{Thacker2013} results agree remarkably well, but at $z \lesssim 0.2$, where FIR background observations predict a weak decline of $\Omega_{\rm dust}$. However, we stress that our $\Omega_{\rm dust}$ derivation is based on an extrapolation of CGM observations down to $z=0$; this may constitute a source of tension. Second, our model predictions broadly reproduce observations within a factor $\sim2$, even better when considering the CGM made of hot+ejected gas (as already highlighted in Sec. \ref{sec:Omega_hot}). Our model does not reproduce any drop in the total dust budget towards $z=0$. However, this drop is evident only for ISM data. \cite{Thacker2013} observations only constrain the whole population of (typically cold) grains responsible for the emission; a hot dust component may be thus missed. On the other hand, CGM observations, which trace the absorption of grains in typically hot environments, do not show any sign of drop up to $z\simeq 0.3$. We thus conclude that, given the available observations, our model reasonably reproduces the cosmic evolution of the total amount of dust. Future observations may better constrain its behaviour, especially for the hot, extra-galactic counterpart.


\section{Summary and conclusions}
\label{sec:conclusion}
In this work, we studied the $z \lesssim 1$ drop of galactic cosmic dust from a semi-analytic model (SAM) perspective. Indeed, integration of the observed DMF in the redshift range $0 < z \lesssim 2.5$ suggests that the total amount of dust in galaxies decreased by a factor $\sim 2-3$ in the last $\sim 8 \, {\rm Gyr}$, albeit the reason for this is not still clear from a theoretical point of view.

We adopted the latest public version of the \textsc{L-Galaxies} SAM \citep{Henriques2020}, where we have incorporated a state-of-the-art dust model and new prescriptions for disc instability. The former is based on the two-size approximation. It considers dust production by type II SNe and AGB stars, grains accretion in molecular clouds, destruction in SNe shocks, sputtering in the hot phase, and shattering of large grains and coagulation of small grains. The original model's disc instability recipe has been updated, taking into account the gaseous disc's contribution to the (in)stability of the whole disc and allowing for star formation and SMBH growth when the instability criterion is met.

This new prescription for disc instability has a non-negligible impact on galaxy evolution. Firstly, it makes the growth of stellar bulges more efficient, leading to a better match with the observed number of local bulge-dominated galaxies. Secondly, it boosts SMBHs growth and thus indirectly enhances quenching, according to the radio-mode feedback scheme adopted in the SAM. We have tuned one of the parameters regulating SMBH growth during instabilities to keep our model predictions in agreement with local SDSS observations of the sSFR distribution. Namely, the adoption of a saturation velocity of the growth efficiency (see Eq. \ref{eq:BHaccinsta}) larger than a factor of $6$ with respect to the one assumed in galaxy mergers, allows us to reproduce the sSFR bimodality observed at ${\rm log \,} M_{\rm stars}/M_\odot \lesssim 10-10.5$, and to suppress the number of highly SF massive galaxies. Some other important quantities (local SMF and HIMF; cosmic SFRD; local BH-bulge mass relation) also agree with observations, as well as the H$_{2}$MF, which, without the new disc instability prescription, is slightly more abundant than the observed one.

The enhanced SMBH growth also shapes the dust properties of our simulated galaxies. While it has just a modest effect on the reproduced DMF at $z>0$, its impact in suppressing the high mass end of the local DMF improves the to match with observations. Moreover, several scaling relations involving dust ($M_{\rm dust}-M_{\rm stars}$; ${\rm DTG}-Z$) turn out to be in reasonable agreement with observations at $0 < z \lesssim 2.5$. Also, the reproduced median $z=0$ small-to-large grains mass ratio reasonably agrees with observations. However, we overproduce the mass of small grains in large $M_{\rm stars}$ (and $Z$) galaxies, suggesting that improved treatment of shattering and coagulation may be needed.\\

Our model allows drawing a few considerations about the behaviour of the cosmic dust density, especially the galactic component, which represents the main focus of this work. We reproduce the observed shape of the galactic cosmic dust density evolution, which is generally a challenging task for cosmological models of galaxies evolution. Specifically, when including the new treatment of disc instabilities, we obtain a more important decrease of the amount of dust in galaxies from $z \sim 1$ to  $z=0$. We found that accretion in molecular clouds is dominant among the processes contributing to the increase of galactic dust. As for processes responsible for the destruction or ejection of galactic dust, astration, destruction in SNe shocks and ejection in winds, all have a non-negligible role. In the fiducial model, their contributions to the total destruction rate of dust in cold gas are, respectively, $\sim 45 \,\%$, $\sim 35 \,\%$, and $\sim 20 \,\%$ at $z=0$. Interestingly, $\Omega_{\rm dust}$ only slightly increases when destruction by SNe shocks is switched off. This happens because the metals returned to the gas phase by this process are quickly re-locked into grains by accretion. Contrarily, switching off astration or dust ejection from cold gas causes an increase by a factor $\sim 2-3 $ of $\Omega_{\rm dust}$. 

The amount of extra-galactic dust in our model continuously grows with time. It broadly reproduces observations but with a steeper redshift evolution, especially when considering both hot gas and the ejected reservoir as CGM. This achievement has been made possible by adopting a sputtering efficiency which is a factor of $10$ lower than the canonical one. This result confirms the claim of hydrodynamic cosmological simulations (\citealt{Gjergo2018}; \citealt{Vogelsberger2019}). The total cosmic dust budget also grows with time, and we argue that more observations of extra-galactic dust, especially at $z\sim0$, are needed to constrain this quantity better.\\

Our main conclusion is that the observed decrease of the galactic dust budget at $z \lesssim 1$ may be reproduced assuming canonical prescriptions for dust production and evolution. This drop is enhanced by our treatment of disc instabilities and SMBH growth, which promotes the suppression of highly SF, dust rich galaxies at $z=0$, providing a better match of the observed local galactic dust abundance. However, although we presented a reasonable model of SMBH growth during disc instabilities, we also speculate that a comparable result could be obtained with different prescriptions. Indeed, many processes may affect the gas content of these galaxies. For instance, SMBH feedback may cause AGN-driven outflows while in the present model, it only brakes gas cooling. 

\section*{Acknowledgements}
\rev{We thank the referee, S. Aoyama, for the comments and suggestions which improved the clarity of our work.}\\
MP would like to thank R. Yates, M. R. Ayromlou and A. P. Vijayan for useful discussions during the \textsc{L-Galaxies}2022 workshop. We also warmly thank A. Bressan, T. Ronconi, M. Behiri, M. Rela$\Tilde{\rm n}$o, and C. Péroux for their constructive comments.\\
This project has received funding from the European Union's Horizon 2020 Research and Innovation Programme under the Marie Sklodowska-Curie grant agreement No 734374, from Consejo Nacional de Investigaciones Cient\'ificas y T\'ecnicas (CONICET) (PIP-2021 11220200102832CO) and from the Ministerio de Ciencia, Tecnología e Innovación (PICT-2020 03690) de la Rep\'ublica Argentina.
Simulations have been carried out at the computing centre of INAF (Italia). We acknowledge the computing centre of INAF-Osservatorio Astronomico di Trieste, under the coordination of the CHIPP project \citep{bertocco2019,Taffoni2020}, for the availability of computing resources and support.

\section*{Data Availability}
The data used for this article will be shared on reasonable request to the corresponding author.


\bibliographystyle{mnras}
\bibliography{bibpeak} 




\appendix


\section{Dust profiles}
\label{app:SLprofile}

\begin{figure}
    \centering

-    \includegraphics[width=0.8\linewidth]{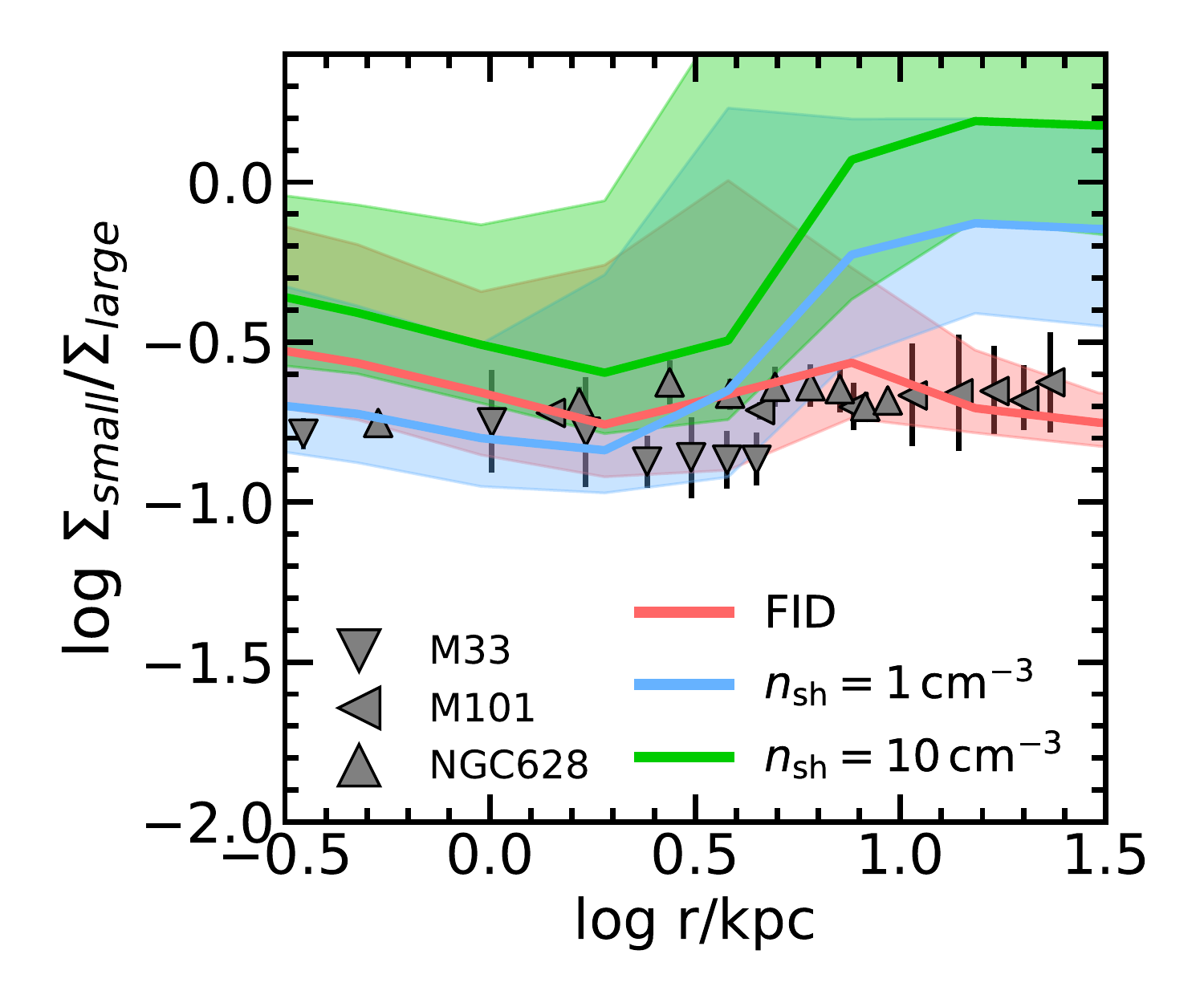} 

    \caption{S-to-L median profile (solid lines) and associated $16-84$th percentile dispersion (shaded regions) for $z=0$ disc (B-to-T $<0.3$) galaxies with $10^9 < M_*/M_\odot < 10^{10.5}$. We show results for our fiducial model, and two models where the number density in Eq. \ref{eq:shat} is assumed to be fixed ($n_{\rm gas}=1 \, {\rm and} \, 10 \, {\rm cm}^{-3}$). Results for three local disc galaxies derived by \citet{Relano2020} are shown as triangles.}
    \label{fig:SL_profile_sha}
    
\end{figure}

The importance of adopting a non-constant $n_{\rm gas}(r)$ in Eq. \ref{eq:shat} (according to Eq. \ref{eq:rhoshat}) is demonstrated by the S-to-L radial profile of galaxies, as shown in Fig \ref{fig:SL_profile_sha}. The S/L profile is too high in the outer regions of the galaxies when assuming $n_{\rm gas}=1$ or $10 \, {\rm cm}^{-3}$. The reason is that these values are excessive for the outer rings. On the contrary, the computation of $n_{\rm gas}$ from $\Sigma_{\rm cold \, gas}$ ensures an almost progressively lower efficiency of shattering with increasing radial distance, as required to match the few available data of local disc galaxies (\citealt{Relano2020}).


\section{The impact of \texorpdfstring{$V_{\rm BH,\, DI}$}{}}
\label{app:BHacc}

In this section, we show the effect of varying the $V_{\rm BH, \, DI}$ parameter during disc instabilities introduced in Eq. \ref{eq:BHaccinsta}. Namely, we show the results obtained assuming a $V_{\rm BH,\, DI}$ during disc instabilities, which is the same as in mergers (Vx1, $700\, {\rm km \, s^{-1}}$), and larger by a factor $6$, $10$ and $20$ (Vx6, Vx10, Vx20). \rev{We also show the results for a run where we adopt a constant $f_{\rm BH, \, unst}=10^{-4}$.} 

\begin{figure*}
    \centering

    \includegraphics[width=1.\linewidth]{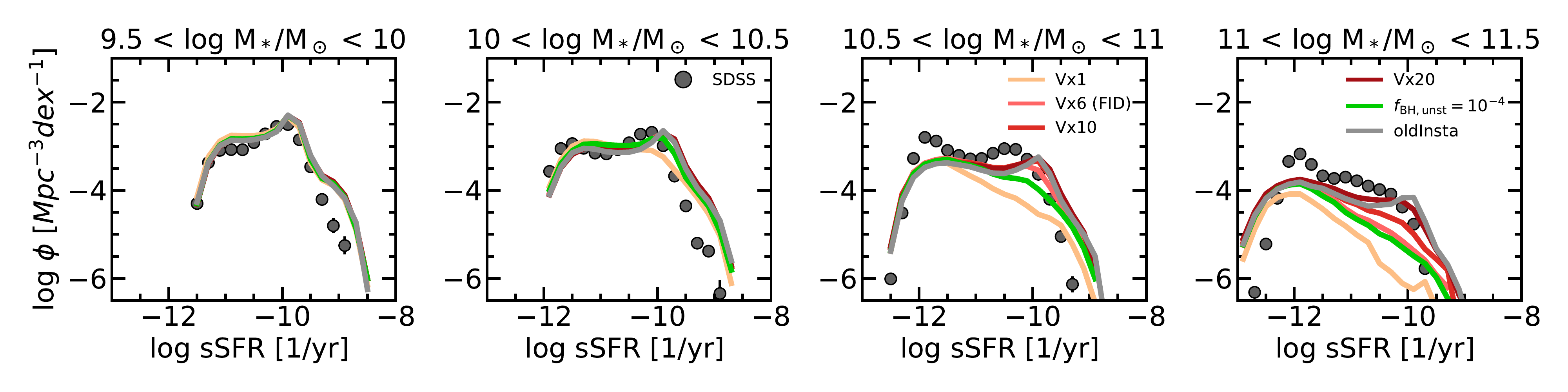} 

    \caption{Same as Fig. \ref{fig:histo-sSFR}, but for models with different $V_{\rm BH,\, DI}$ parameters during disc instabilities (eq. \ref{eq:BHaccinsta}). Namely, we show the results obtained assuming the same $V_{\rm BH}=700 \, {\rm km\,s^{-1}}$ in both mergers and instabilities (Vx1), and a $V_{\rm BH, \, DI}$ larger by a factor $6$, $10$ and $20$ (Vx6, Vx10, Vx20) during disc instabilities. The Vx6 model is the fiducial one adopted in this work (\textit{FID}). \rev{We also show the results obtained assuming a constant $f_{\rm BH, unst}=10^{-4}$ (green line).} Results for the model without the new implementation of disc instabilities are shown for comparison (\textit{oldInsta}, gray line).}
    \label{fig:BHacc:sSFR}
    
\end{figure*}

The impact of these choices on the $z=0$ sSFR distribution in different stellar mass ranges is shown in Fig. \ref{fig:BHacc:sSFR} (same as Fig. \ref{fig:histo-sSFR}). This quantity is particularly convenient for investigating the impact of the SMBH growth on the quenching of galaxies with different masses. For example, when adopting the same $V_{\rm BH}$ in mergers and instabilities (Vx1) we observe a too large quenched galaxies fraction with respect to the star forming counterparts when compared to SDSS data. Adopting a larger $V_{\rm BH,\, DI}$ helps to mitigate the SMBH growth and their consequent effect on quenching. In particular, we find that $V_{\rm BH, DI} = 6 \times V_{\rm BH, mergers}$ (Vx6) keeps our model predictions in good agreement with SDSS data for ${\rm log} \, M_*/M_\odot < 11$. Moreover, in the largest stellar mass bin ($11 < {\rm log}\, M_*/M_\odot < 11.5$), the Vx6 model still suppresses the formation of high sSFR galaxies, which are instead obtained in the \textit{oldInsta} model. Similar conclusions may be drawn for the Vx10 model, which unsurprisingly produces a larger number (with respect to the Vx6 model) of high sSFR galaxies in the most massive stellar mass bin. Increasing $V_{\rm BH, \, DI}$ up to $20$ times the value adopted in mergers (Vx20), the sSFR distribution mostly resembles what is obtained in the \textit{oldInsta} model. 

\rev{As for the $f_{\rm BH, \, unst}=10^{-4}$ model, it reproduces quite well the results obtained with our fiducial model broadly, but for a slightly worse agreement with SDSS data in the $10.5 < {\rm log} \, M_*/M_\odot < 11$ bin. This difference has to be ascribed to the $V_{\rm vir}$ dependence, which is adopted in our fiducial model (Eq. \ref{eq:BHaccinsta}) and ensures SMBH growth to be more effective in more massive systems.}

\begin{figure}
    \centering

    \includegraphics[width=1.\linewidth]{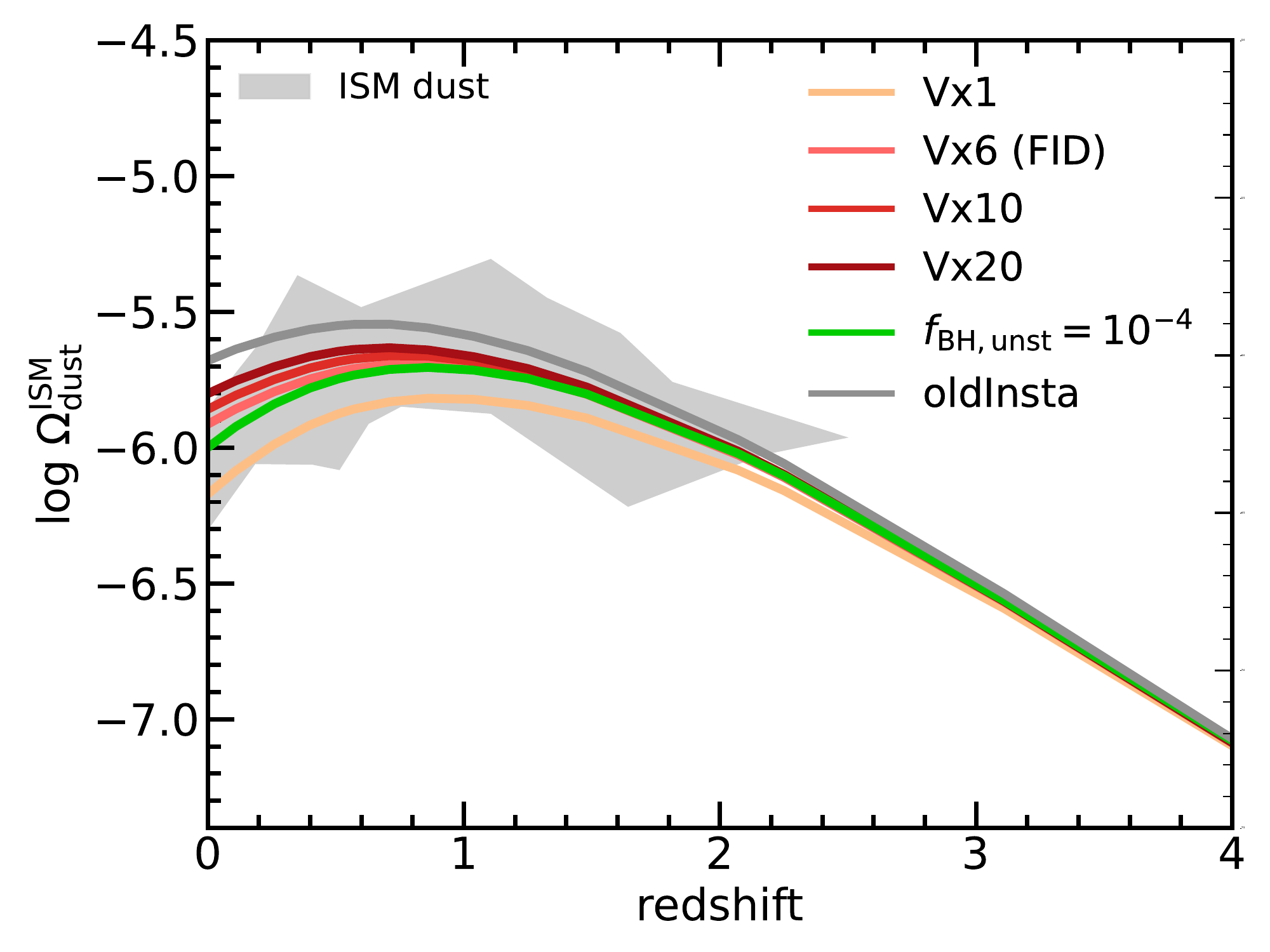} 

    \caption{Cosmic evolution of galactic dust obtained by models with different $V_{\rm BH, \, DI}$ (Eq. \ref{eq:BHaccinsta}). We show the results obtained assuming the same $V_{\rm BH}=700 \, {\rm km\,s^{-1}}$ in both mergers and instabilities (Vx1), and a $V_{\rm BH, \, DI}$ larger by a factor $6$, $10$ and $20$ (Vx6, Vx10, Vx20) during disc instabilities. The Vx6 model is the fiducial one adopted in this work (\textit{FID}). \rev{We also show the results obtained assuming a constant $f_{\rm BH, unst}=10^{-4}$ (green line).} Results for the model without the new implementation of disc instabilities is shown for comparison (\textit{oldInsta}, gray line). We also show the compilation of observations introduced in Sec. \ref{sec:Omega_gal} (gray shaded region).}
    \label{fig:OmegaVBH}
    
\end{figure}

Finally, we show in Fig. \ref{fig:OmegaVBH} the impact of varying $V_{\rm BH, \, DI}$ on the cosmic evolution of the galactic $\Omega^{\rm ISM}_{\rm dust}$. As expected from the above discussion, the larger is $V_{\rm BH, \, DI}$, the lower is the suppression of highly SF galaxies, and thus the higher is the $\Omega^{\rm ISM}_{\rm dust}$ normalization. \rev{Similarly, the $f_{\rm BH, \, unst}=10^{-4}$ model produces a somewhat stronger $\Omega^{\rm ISM}_{\rm dust}$ drop than the fiducial model, as a result of its slightly stronger SF suppression.}\\

\section{Galactic sizes}
\label{app:sizes}
Here we briefly discuss the size of the galaxies within our SAM. Although not significantly relevant for the goals of this work, they are affected by the disc instability process, which transfers material from the disc to the bulge, modifying their mass distributions and thus sizes.
We take inspiration from \cite{Irodotou2019} for determining the size of bulges formed/grown as a result of both mergers and disc instabilities. 
\begin{itemize}
    \item \textbf{Major Mergers} \\
    In major mergers, all the pre-existing stars and those formed during the merger, following the collisional starburst formulation by \cite{Somerville2001}, are added to the bulge of the descendant galaxy. Its half-mass radius $R_{\rm fin, \, bulge}$ is computed from energy conservation arguments (see also \citealt{Tonini2016}):
    \begin{equation}
        \frac{(M_{*,1}+M_{*,2}+M_{*,\rm{burst}})^2}{R_{\rm fin,\, bulge}} = \left(\frac{M^2_1}{R_1}+\frac{M^2_2}{R_2}\right)\left(1+k_{\rm rad}\right) + \left( \frac{M_1M_2}{R_1+R_2} \right). 
    \end{equation}
    In the above formula, $M_{i}$ are the total baryonic mass of the progenitors, $M_{*,i}$ their total stellar mass, $R_{i}$ their stellar half-mass radii, and $M_{\rm *,\, burst}$ the stellar produced in the burst. The term $k_{\rm rad}$ includes energy dissipation by gas due to radiative losses, and it is computed as:
    \begin{equation}
    \label{eq:radloss}
        k_{\rm rad} = C_{\rm rad} \left(\frac{M_{\rm gas,1}+M_{\rm gas,2}}{M_1 + M_2}\right),
    \end{equation}    
    being $M_{\rm gas,\, {i}}$ the gas mass of the progenitors, and $C_{\rm rad}$ is a parameter of order $\simeq 1$ \citep{Covington2008} quantifying the efficiency of radiative losses. Here set to $=1$. 
    
    \item \textbf{Minor Mergers}\\
    Minor mergers are modelled assuming that the disc of the larger progenitor survives and will host the smaller progenitor's cold gas and the stars formed during the merger-driven starburst. Instead, the larger galaxy's bulge accretes the second progenitor's stars. Its size $R_{\rm fin,\, bulge}$ is computed according to:
    \begin{equation}
        \frac{(M_{\rm bulge,1}+M_{*,2})^2}{R_{\rm fin,\, bulge}} = \left(\frac{M^2_{\rm bulge, 1}}{R_{\rm bulge, 1}}+\frac{M^2_{*,2}}{R_2}\right)\left(1+k_{\rm rad}\right) + \left( \frac{M_{\rm bulge, 1}M_{*,2}}{R_{\rm bulge, 1}+R_2} \right). 
    \end{equation}
    In the above expression, $M_{\rm bulge, 1}$ and $R_{\rm bulge, 1}$ are the mass and half-mass radius of the bulge of the first progenitor \textit{before} the merger, $M_{\rm *, 2}$ and $R_{\rm *, 2}$ the mass and half-mass radius of the stellar component of the smaller progenitor. The $k_{\rm rad}$ term is computed as in Eq. \ref{eq:radloss}.
    
    \item \textbf{Disc instabilities}\\
    If the galaxy already possesses a bulge with mass $M_{\rm in}$ and half-mass radius $R_{\rm in}$, the bulge size $R_{\rm fin}$ after a disc instability event leading to a final bulge mass $M_{\rm fin}$ is given by:   
    \begin{equation}
    \frac{M^2_{\rm fin}}{R_{\rm fin}} = \frac{M^2_{\rm in}}{R_{\rm in}} + \frac{M^2_{\rm unst}}{R_{\rm unst}} + \frac{\alpha}{C} \frac{M_{\rm in}M_{\rm unst}}{R_{\rm in}+R_{\rm unst}},
    \label{eq:bulgesizeDI}
    \end{equation}
    with $\frac{\alpha}{C}=1.5$ ($C$ is a structural parameter and $\alpha$ regulates the energy dissipation; \citealt{Boylan-Kolchin2005}). In the above expression, $M_{\rm unst}$ is the mass of the material transferred from disc to bulge as a result of the instability, and $R_{\rm unst}$ its half-mass radius\footnote{Note that, since the unstable material is already in rings, we can trivially compute the half mass radius of the unstable mass.}.\\
    When a galaxy does not possess a bulge before a disc instability event ($M_{\rm in}=0$), we simply assume $R_{\rm fin}=R_{\rm unst}$.\\

\end{itemize}

\begin{figure*}
    \centering

    \includegraphics[width=0.8\linewidth]{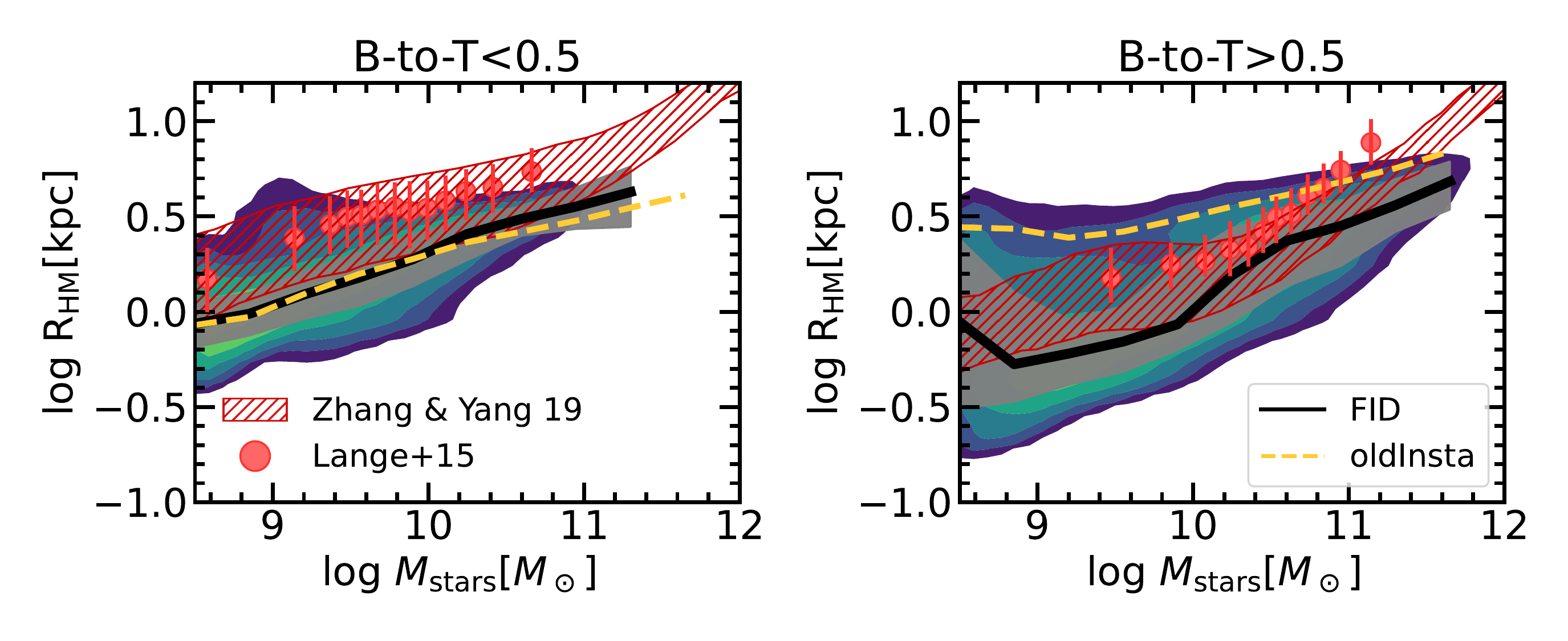} 

    \caption{Stellar half-mass radius as a function of stellar mass for disc-dominated (\textit{left panel}, B-to-T$<0.5$) and bulge-dominated (\textit{right panel}, B-to-T$>0.5$). The black line and gray shaded area represent the median and $16-84$th percentiles dispersion of our fiducial model results, while log-spaced density contours are shown in background. The median relation obtained by the \textit{oldInsta} model is shown as a gold dashed line. We compare with observations of \citet{Lange2015} (red circles; we use their classification of disc and bulge dominated galaxies according to the Sérsic index) and \citet{Zhang2019} (hashed red regions, corresponding to the $16-84$th percentiles of the observed dispersion).}
    
    \label{fig:sizes}
    
\end{figure*}

In Fig. \ref{fig:sizes} we show the stellar half-mass radius $R_{\rm HM}$ for disc dominated (B-to-T $< 0.5$) and spheroid dominated (B-to-T $>0.5$) galaxies. We compare with observations by \cite{Lange2015} and \cite*{Zhang2019}. In general, our trends are in broad agreement with the shown data. Nevertheless, our fiducial model predicts a median stellar half-mass radius of disc-dominated galaxies, a factor $\sim 2$ lower than observations. We point out that this result is only mildly affected by our implementation of disc instability since a similar behaviour is observed for the \textit{oldInsta} model. As for bulge-dominated galaxies, the new computation of bulge sizes introduced in this section improves the performance of our fiducial model in predicting lower stellar half-mass radii, more in line with observations.\\


\section{BH-BULGE MASS RELATION}
\label{app:sanity2}

\begin{figure}
    \centering

    \includegraphics[width=0.8\linewidth]{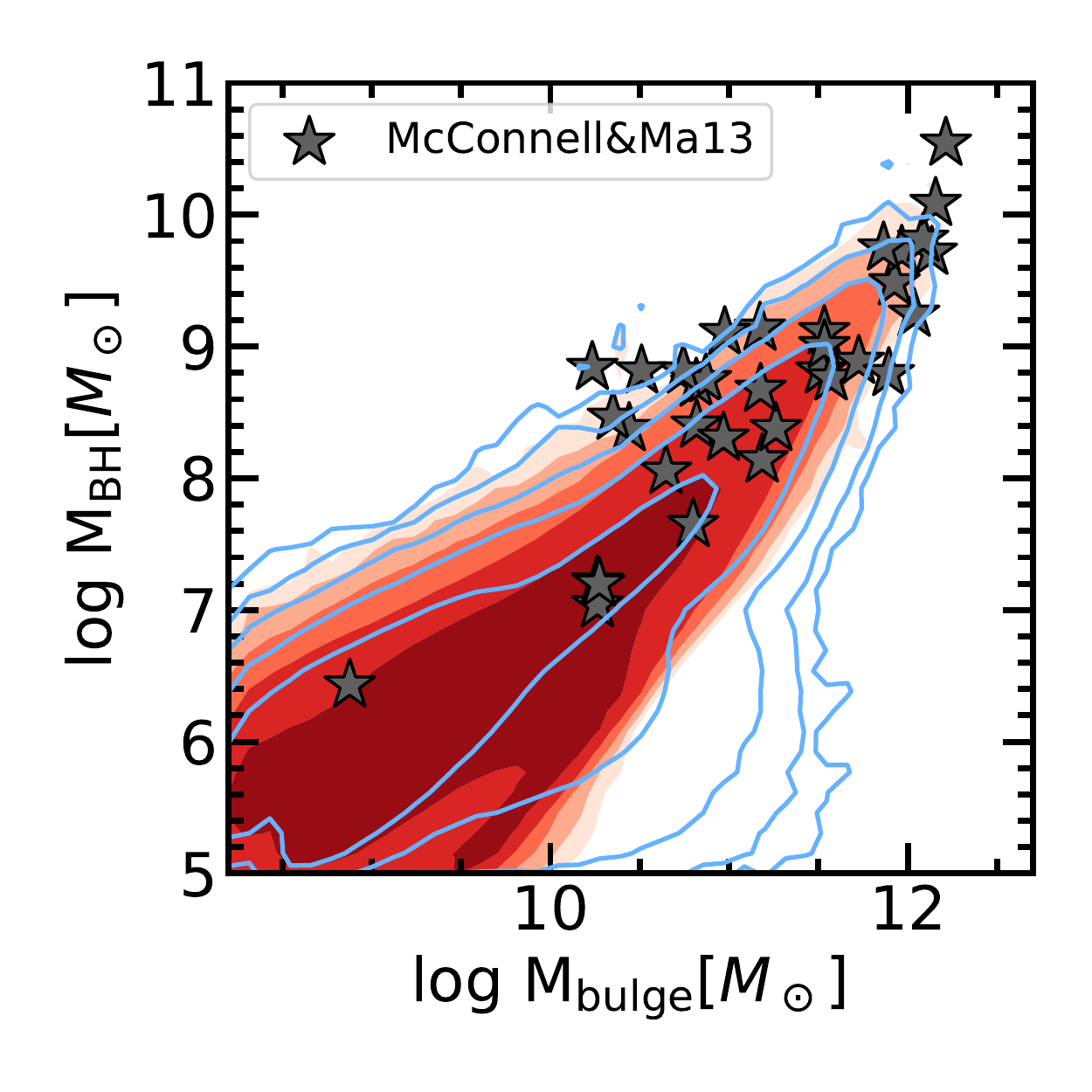} 

    \caption{BH-bulge mass relation at $z=0$ for our fiducial model (log spaced filled red contours of the density of galaxies in this plane) and for the \textit{oldInsta} model (blue contours) compared with observations by \citet{McConnell2013}.}
    \label{fig:BHbulge}
    
\end{figure}

The BH-bulge mass relation at $z=0$ is shown in Fig. \ref{fig:BHbulge} for both the \textit{FID} and \textit{oldInsta} model. The new implementation of BH growth during disc instabilities produces slightly larger BH masses, but still in agreement with most of the observations by \cite{McConnell2013}.\\

\bsp	
\label{lastpage}
\end{document}